\documentclass{aa} 

\usepackage{graphicx}
\usepackage{txfonts}

\usepackage[english]{babel}
\usepackage{amsmath}
\usepackage{amssymb}    
\usepackage{multicol}        
\usepackage{bm}         
\usepackage{pdflscape}  
\usepackage{flafter}
\usepackage{wrapfig}

\usepackage[export]{adjustbox}
\usepackage{float}
\usepackage{enumitem}
\usepackage{tabularx}
\usepackage[normalem]{ulem}
\usepackage{subfig}
\usepackage{xcolor}

\begin{document} 
   
   \title{Discovering planets with PLATO: Comparison of algorithms for stellar activity filtering\footnote{Tables of the light curves are only available in electronic form at the CDS via anonymous ftp to cdsarc.cds.unistra.fr (130.79.128.5) or via https://cdsarc.cds.unistra.fr/cgi-bin/qcat?J/A+A/}}

   \author{G.~Canocchi\inst{\ref{inst1}, \ref{inst2}} \and
          L.~Malavolta\inst{\ref{inst2}, \ref{inst3}} \and
          I.~Pagano\inst{\ref{inst4}} \and 
          O.~Barragán\inst{\ref{inst5}} \and 
          G.~Piotto\inst{\ref{inst2}, \ref{inst3}} \and
          S.~Aigrain\inst{\ref{inst5}} \and 
          S.~Desidera\inst{\ref{inst3}} \and 
          S.~Grziwa\inst{\ref{inst6}} \and
          J.~Cabrera\inst{\ref{inst7}} \and
          H.~Rauer\inst{\ref{inst7}}
          }

    \institute{Department of Astronomy, Stockholm University, AlbaNova University Center, SE-106 91 Stockholm, Sweden\\ \email{gloria.canocchi@astro.su.se}\label{inst1} \and 
    Dipartimento di Fisica e Astronomia ``Galileo Galilei'' – Universit\`{a} di Padova, Vicolo dell’Osservatorio 3, 35122 -- Padova, Italy\label{inst2}\and 
    INAF -- Osservatorio Astronomico di Padova, Vicolo dell'Osservatorio 5, 35122 -- Padova, Italy\label{inst3} \and 
    INAF -- Osservatorio Astrofisico di Catania, via S. Sofia 78, 95123 -- Catania, Italy\label{inst4} \and 
    Sub-department of Astrophysics, Department of Physics, University of Oxford, Oxford, OX1 3RH, UK\label{inst5} \and
    Rheinisches Institut für Umweltforschung an der Universiät zu Köln, Aachener Straße 209, D-50931 Köln, Germany \label{inst6} \and 
    Institute of Planetary Research, German Aerospace Center, Rutherfordstrasse 2, D-12489 Berlin, Germany \label{inst7}
    }

  \abstract  
   {To date, stellar activity is one of the main limitations in detecting small exoplanets via the transit photometry technique. Since this activity is enhanced in young stars, traditional filtering algorithms may severely underperform in attempting to detect such exoplanets, with shallow transits often obscured by the photometric modulation of the light curve.}
   {This paper aims to compare the relative performances of four algorithms developed by independent research groups specifically for the filtering of activity in the light curves of young active stars, prior to the search for planetary transit signals: \textit{Notch and LOCoR (N\&L)}, \textit{Young Stars Detrending (YSD)}, \textit{K2 Systematics Correction (K2SC),} and \textit{VARLET}.  Our  comparison also includes the two best-performing algorithms implemented in the \texttt{W{\=o}tan} package: Tukey's \textit{biweight} and \textit{Huber spline} algorithms.}
   {For this purpose, we performed a series of injection-retrieval tests of planetary transits of different types, from Jupiter down to Earth-sized planets, moving both on circular and eccentric orbits. These experiments were carried out over a set of 100 realistically simulated light curves of both quiet and active solar-like stars (i.e., F and G types) that will be observed by the ESA \textit{Planetary Transits and Oscillations of stars} (PLATO) space telescope, starting 2026. }
   {From the experiments for transit detections, we found that \textit{N\&L} is the best choice in many cases, since it misses the lowest number of transits. However, this algorithm  is shown to underperform when the planetary orbital period closely matches the stellar rotation period, especially in the case of small planets for which the \textit{biweight} and \textit{VARLET} algorithms work better. Moreover, for light curves with a large number of data-points, the combined results of two algorithms, \textit{YSD} and \textit{Huber spline}, yield the highest recovery percentage. Filtering algorithms allow us to obtain a very precise estimate of the orbital period and the mid-transit time of the detected planets, while the planet-to-star radius is underestimated most of the time, especially in cases of grazing transits or eccentric orbits. A refined filtering that takes into account the presence of the planet is thus compulsory for proper planetary characterization analyses. }
   {}

\titlerunning{Discovering planets with PLATO}
\authorrunning{Canocchi G. et al.}
   \maketitle\footnote{Tables of the light curves are only available in electronic form
at the CDS via anonymous ftp to cdsarc.cds.unistra.fr (130.79.128.5)
or via https://cdsarc.cds.unistra.fr/cgi-bin/qcat?J/A+A/}

\section{Introduction}\label{sec: intro}
From an observational point of view, stellar activity poses a challenge to detections of planet around stars different from the Sun and to precise measurements of their parameters, especially with regard to the radial velocity and the transit photometry techniques. In stellar light curves (LCs), the out-of-transit flux is not typically flat but, instead, it is characterized by amplitude variations that are due to the intrinsic stellar activity, which is particularly enhanced in young stars (i.e., $<$1 Gyr). The latter do indeed show flux variations on many different timescales, with amplitudes up to 700\% in a temporal range of 1-10 days (\citealt{Cody2017}; \citealt{CodyHillenbrand2018}). These fluctuations in brightness are caused by several phenomena occurring on the stellar surface, primarily spots and faculae, that is, colder or brighter regions on the stellar surface, which can remove or add flux, depending on their location when the planet is transiting, thus affecting both the transit depth and shape, as well as the overall LC.
Another source of noise for transit parameter determination are stellar oscillations, granulation, and flares, which contribute to increase the flux variability, consequently affecting planet detection. The effect of stellar activity on transit is more remarkable  the smaller the planet is. In particular, young stars are characterized by fast rotation and complex brightness modulation, which makes them challenging targets for exoplanet research.\\ 
Despite these difficulties, observing young stars is important for multiple reasons: these young exoplanets can provide hints about the first stages of planet formation and evolution, helping us to put some constraints on current formation models. Understanding the mechanisms occurring in the early evolutionary phase of planet formation -- such as migration and ionizing radiation from the host star (\citealt{Baraffe2003}), as well dynamical interactions with the protoplanetary disk, the other bodies in the system (e.g., the Kozai-Lidov mechanism, \citealt{Shevchenko2017}), or with nearby stars (\citealt{IdaLin2010}) -- is fundamental in order to better interpret the final distribution of extrasolar planet population. Indeed, all these processes contribute to change both orbital and planetary parameters, such as mass, eccentricity, and distance from the host star, especially in the first phase of planet formation. \\ \\
The aforementioned stellar activity in young stars can be of roughly on the same order of magnitude  as that of the transit signal, in terms of both period and flux variation (\citealt{Armstrong2015}), thus masking the true planetary signals or producing false positives (\citealt{Rodenbeck2018}). 
This is the reason why most of the transit surveys are focused on older, photometrically quieter stars, exhibiting only small variations over long time scales (\citealt{Rowe2014}). Nevertheless, when looking for Earth-sized planets around solar-like stars, even this low variability of the stellar flux over long time scales can become problematic for the transit detection. Therefore, many filtering algorithms have been developed in order to correct for stellar activity in photometric LCs.\\
According to \citet{Hippke2019}, in planet injection-retrieval experiments, these methods perform very well in quiet stars, reaching a detection efficiency of nearly 100\%. However, this percentage decreases down to $\sim 43.8\%$ when considering a sample of young, extremely active stars observed with the \textit{Transiting Exoplanet Survey Satellite} (\textit{TESS}, \citealt{TESSRicker}), meaning that less than half of the injected transit signals are recovered after detrending the highly variable LCs produced by these young stars. In their tests, Hippke et al. (2019) considered planets with a radius equal to half the Jupiter radius, hence, the recovery fraction of transit signals is expected to decrease even more when considering small planets such as mini-Neptunes and Super-Earths. This proves how challenging is searching for young exoplanets around young active stars. \\ \\

Several algorithms have been proposed to specifically filter activity in the LCs of young stars in order to search for planetary transit signals, leading to the discovery of young exoplanets (e.g., \citealt{Mann2016}, \citealt{Newton2019}, \citealt{Benatti2019}, \citealt{Nardiello2019}, \citealt{Thao2020}).
However, many more young exoplanets are needed for building reliable statistics.\\  
The main purpose of this work is to identify and compare algorithms specifically developed for the filtering of stellar activity in young stellar LCs, in order to test their effectiveness prior to the search of transit events. We performed tests on realistically simulated LCs of both quiet and active solar-like stars, as these are the types that will be observed by the forthcoming ESA mission \textit{PLAnetary Transits and Oscillations of stars} (\textrm{PLATO}, \citealt{PLATOref}). In particular, we selected four custom-built algorithms: \texttt{Notch and LOCoR}\footnote{https://github.com/arizzuto/Notch\_and\_LOCoR} (N\&L, \citealt{Rizzuto2017}), \texttt{Young Stars Detrending}\footnote{https://github.com/mbattley/YSD} (YSD, \citealt{YSD}), \texttt{K2 Systematics Correction}\footnote{https://github.com/OxES/k2sc} (K2SC, \citealt{Aigrain2016}) and \texttt{VARLET and PHALET} (\citealt{VarletPhalet}).\\ These have been recently successfully applied by independent groups on \textrm{Kepler} (\citealt{Keplerref}), \textrm{K2} (\citealt{K2ref}) or \textit{TESS} LCs, specifically for the search of planets around young, active stars.
Along with the above-mentioned algorithms, we also included in the comparison two of the algorithms selected as the best performing ones for less active stars according to Hippke et al. (2019), namely, "Tukey's biweight" and "Huber spline," as implemented in the \texttt{W{\=o}tan}\footnote{https://github.com/hippke/wotan} Python package. This paper is organized as follows. In Sect. \ref{sec:PLATO LCs}, the \textrm{PLATO} mission is introduced, along with the description of the samples of \textrm{PLATO} simulated LCs and the criteria for transit detection. A brief summary of the algorithms tested by Hippke et al. (2019) (i.e., general-purpose algorithms) and the analysis of their performance on a series of different transit signals injected into \textrm{PLATO} LCs is presented in Sect. \ref{sec: Hippke algorithms}. In Sect. \ref{sec: new algorithms} the custom-built algorithms mentioned above are introduced. Then, in Sect. \ref{sec: tests}, several injection-retrieval tests are performed on both individual quarters (Sect. \ref{sec: individual_quarters}) and entire LCs (Sect. \ref{sec: entire_LC}).
Moreover, we  investigate some peculiar cases: eccentric hot Jupiters (Sect. \ref{sec: HJ}) and planets with an orbital period equal to the rotation period of their host star (Sect. \ref{sec: Porb=Prot}). Finally, in Sect. \ref{sec: conclusions}, we discuss the main results and future perspectives.\\ 

\section{PLATO simulated light curves and transit detection}
\label{sec:PLATO LCs}
\textrm{PLATO}\footnote{\url{https://www.cosmos.esa.int/web/plato}} (\citealt{PLATOref}) is the third medium-class mission in ESA's Cosmic Vision program. Scheduled for launch in 2026, \textrm{PLATO} has the main aim of detecting and characterizing, in terms of density and age, Earth-like planets around solar-like stars up to the habitable zone.
\textrm{PLATO} will carry out high-precision photometric observations on a large sample of stars in order to characterize both the exoplanets and their host stars, respectively, through transit detection and asteroseismology.
Since \textrm{PLATO} is designed to observe a broad field of view (FoV)\footnote{\textrm{PLATO} should cover between 10\% and 50\% of the whole sky, depending on the adopted strategy.}(\citealt{Nascimbeni2022}), similarly to the NASA all-sky survey mission \textit{TESS}, a multi-telescope approach was adopted. Indeed, \textrm{PLATO} is equipped with 24 "normal" cameras or N-CAM (i.e., 25 s readout cadence), organized into four groups of six cameras, each with its own CCD-based focal plane array, and two "fast" cameras or F-CAM (i.e., 2.5 s readout cadence) dedicated to the observation of bright stars through the use of two bandpass filters (one per camera) acting as fine guidance sensors. All the cameras are refracting telescopes with an aperture diameter of about 120 mm, pointing at different parts of the sky (with some overlapping areas), covering a total FoV of more than 2200 deg$^2$ \citep{Pertenais2021,Magrin2020}.
Each camera is equipped with four charge-coupled devices (i.e., 104 CCDs in total) with 4510 $\times$ 4510 pixels of size 18 $\mu$m (edge length). 
In preparation of the future \textrm{PLATO} space mission, several research groups implemented simulated LCs as those expected to be obtained by the \textrm{PLATO} telescope. According to the current observation strategy, \textrm{PLATO} will point the same portion of the sky for at least two years, but slightly adjusting the spacecraft position every three months (i.e., every quarter), rotating 90$\mathrm{^o}$ around the line-of-sight. Hence, considering the total time-series datasets, between one quarter and the next,  a jump in the LC is expected, due to the fact that the pixel position on the camera where a given target is observed can change between a quarter and another. 
\subsection{Simulated light curves}
For our analysis, we employed two samples of 100 simulated LCs provided by the \textit{Light curve Stitching Working Group} (LSWG), which is investigating the best method to model the flux variations between quarters of \textrm{PLATO} LCs. Each LC encompasses eight quarters with a duration of $\sim$ 88 days and gaps between quarters of 1-3 days, covering 2 years in total.
The LCs were originally produced at a sampling of 25 s. Since such an high cadence was not needed for the transit detection, in order to save computational time and keep the file size manageable, the curves were binned to 600 s cadence. 

The LSWG produced 100 LCs simulating \textrm{PLATO}-like data of bright stars ($8 \leq V \leq 11$ mag)\footnote{One merged LC is provided, even if the star is observed by 24 cameras.}
with a photometric amplitude variability in the range 0.08--1.71\%, representative of \textrm{PLATO}-like systematics, as obtained by the \texttt{PLATO Solar-like Light curve Simulator} (PSLS) 1.2\footnote{https://sites.lesia.obspm.fr/psls/} (\citealt{Samadi2019}).
We note that for this particular set of simulations we did not use any realistic pointing direction, that is, the noise values are representative of what \textrm{PLATO} will observe, but there is no correlation with the distribution of stars in the sky.\\ As the mission is still under development, many technical details such as the noise levels, the duration of the runs (in this work: 88 days), and the duration of the gaps between the quarters (in this work: between 1-3 days) are based on requirements or realistic assumptions, while real in-flight performance might be different.
The instrumental systematics and random noise for the noise-only LCs were simulated assuming point spread function (PSF) photometry (as will be done on the ground for the bright stars in the P1 sample). The final parameter that controls the noise in PSLS is the stellar magnitude, which is reported in Table \ref{tab: LC param-quiet} and \ref{tab: LC param-active} in Appendix \ref{app: stars}, along with the full set of input parameters for the LCs used in this work.\\
For the activity signals, we required a physically motivated model where individual parameters such as the rotation rate and the spot coverage could be controlled by the user. This is only partially accomplished by the stellar activity component in PSLS, which consists of a Lorentzian in Fourier space with no explicit periodicity (see Sect. 5.2 of \citealt{Samadi2019} for more details). Hence, we switched off the variability components of PSLS and manually inserted activity signals and oscillations  afterwards.
Therefore, onto these noise-only LCs, we superimposed the activity signals of solar-like "quiet" stars with $P_\mathrm{rot} \sim 1-50$ days  extracted from LCs simulated in \citet{Aigrain2015}, which included Sun-like butterfly patterns, activity cycles, spot evolution (i.e., emergence and decay)\footnote{In 
\citet{Aigrain2015} they explicitly simulate the effect of each spot on the LC using a simple spot model.}, and differential rotation, on a temporal range of 1000 days and representative of the \textit{Kepler} data of solar-like stars. We refer to this dataset as the "quiet sample."\\ 
The "active sample" required to test the different algorithms was based again on the same LCs produced by the LSWG, this time with the addition of a strong quasi-periodic activity signal with a rotation period shorter than 10 days\footnote{In fact, rotational velocity generally decreases with age (i.e., rotation period increases) as well as activity (\citealt{Salabert2016}), depending also on the color index (\citealt{Mamajek2008}).}, again extracted from the simulations performed by \citet{Aigrain2015}.
Two examples, representing the LCs of a quiet and an active star respectively, as simulated by the LSWG are shown in Fig. \ref{fig: PLATOLC_example} (top) and (bottom), respectively. 

\begin{figure}[H]
   \centering
   \includegraphics[width=9cm]{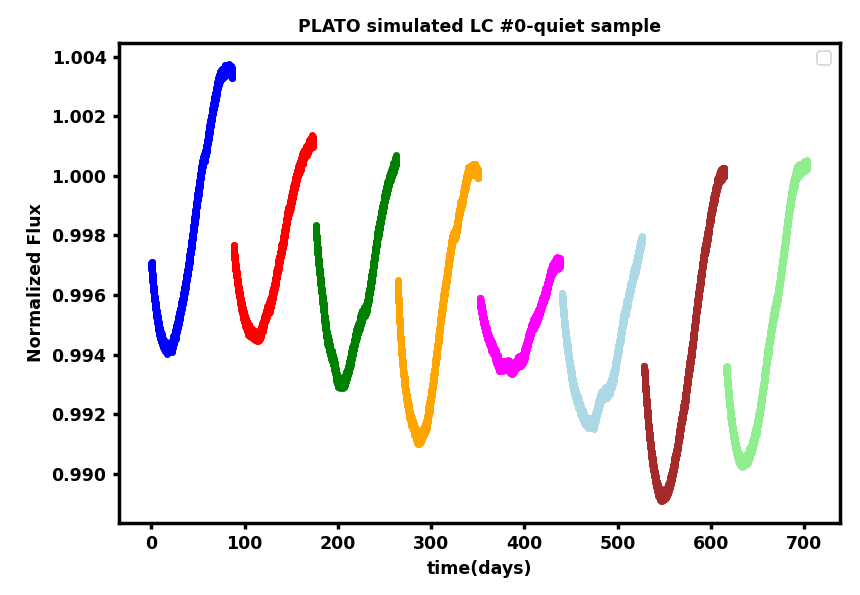}
   \centering
 \end{figure}\vspace{-10pt}
 \begin{figure}[H]
 \centering
   \includegraphics[width=9cm]{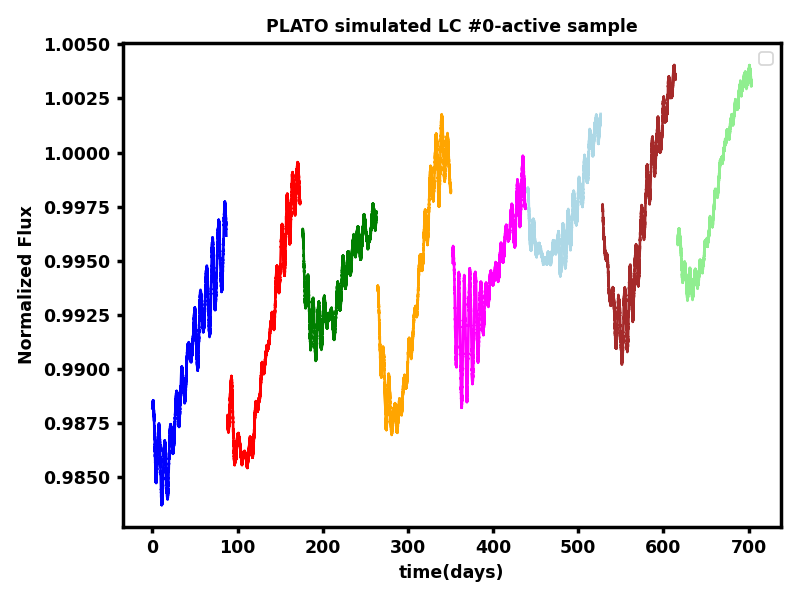}
      \caption{\textrm{PLATO} simulated LC \#0 of the sample of quiet stars (top) and active stars (bottom). Different colors represent different quarters. The systematics in the simulated LCs represent  a worst-case scenario, as their correction will be addressed in the final \textrm{PLATO} data products.}
         \label{fig: PLATOLC_example}
   \end{figure}\vspace{-10pt}

Finally, we did not attempt to remove the instrumental long-term systematics which are very obvious in the top panel of Fig. \ref{fig: PLATOLC_example}, as the algorithms that will address those residuals in the \textrm{PLATO} pipeline are still under development. The LCs employed in this analysis should not be considered as representative of \textrm{PLATO} final data products, which indeed will be corrected for instrumental systematics\footnote{See Chapter 8.1 PLATO data products of the \textrm {PLATO} Definition Study Report, \url{https://sci.esa.int/s/8rPyPew}} but should rather be considered a worst-case scenario in which intermediate data products are employed for the analysis.

\subsection{Transit modeling} 
Transits were modeled using the \texttt{batman}\footnote{https://github.com/lkreidberg/batman} Python package (\citealt{Kreidberg2015}), which is a Python implementation of the transit model from \citet{Mandel2002}. Since we wanted to measure the performances of the algorithms on different populations of exoplanets, the orbital parameters and planetary properties were varied from case to case, and they are described along the paper accordingly. The injection was performed by simply multiplying the stellar LC by the transit model (normalized to unity outside the transit), that is, spot-crossing events and other effects altering the shape of the transit were not included. The model did not include other planet-related effects other than the transit, for instance, phase curve variations, secondary eclipses, Doppler boosting, and ellipsoidal variations were not included. 
After the transit injection, these LCs were detrended with selected filtering algorithms (see Sects. \ref{sec: Hippke algorithms} and \ref{sec: new algorithms}). The transit detection on the resulting filtered flux was then performed by the \textsc{Transit Least Squares} (TLS) algorithm \citep{TLS2019}, which has shown a higher detection efficiency compared to other transit detection algorithms, such as \textsc{Box-fitting Least Squares} (BLS) \citep{BLS}. 
The main improvement of TLS with respect to BLS is that it employs a real planetary transit shaped model when searching for transit signals in the phase-folded corrected LC and not simply a less realistic box-like shape. The TLS default template of the transit curve used for transit detection in this work, assumes circular orbits (i.e., a null eccentricity), with an impact parameter of 0 (i.e., an inclination of 90 degrees).
Unfortunately, we realized at a later stage that the filtering algorithms may modify the transit signal to a point that there is no real advantage in using a more accurate shape for the transit; thus, in the future, we advise to use more than one detection technique in order to maximize the scientific return.

TLS computes the Signal Detection Efficiency (SDE) as the detection statistics for finding the strongest signal in the periodogram consistent with a planetary transit. The SDE distribution as a function of the orbital period, is obtained  in the same way as in \citet{BLS}:
\begin{equation}
    {\rm SDE}(P_{\rm orb})= \frac{{\rm SR}_{\rm peak}-<{\rm SR}(P_{\rm orb})>}{\sigma({\rm SR}(P_{\rm orb}))}
    \label{eq: SDE}
,\end{equation}
where ${\rm SR}(P_{\rm orb})$ is the signal residual between the model and the data, computed from the distribution of minimum $\chi^{(2)}$. In particular: $<{\rm SR}(P_{\rm orb})>$ is the arithmetic mean, $\sigma({\rm SR}(P_{\rm orb}))$ is the standard deviation, and $\rm SR_{peak}$ is the peak value, which (by definition) is equal to 1. Once that the SDE distribution has been computed, the period corresponding to the highest SDE value is assumed as the orbital period of the transit signal. In our analysis, the planet was considered to have been recovered if the detected period matched the injected one within 1\% . Moreover, for a more realistic transit search, a detection SDE threshold of 7 was applied, as done by \cite{HippkeHeller2019}. This means that all signals with SDE below this threshold were not considered as correctly identified, even if the recovered orbital period matched the injected one within 1\%. We also made use of the SDE of the recovered period as a metric for the comparison of different algorithms. We are aware that the SDE threshold should depend on the properties of the light curves and that a single value may not be appropriate (see for example \citealt{EXOTRANS}), but we believe it still represents a reasonable choice given the goals of this work. We stress again the fact that our choice is not representative of the final implementation of the \textrm{PLATO} pipeline, and it is only valid in the framework of this specific work. Finally, we note that we did not produce any comparison sample, that is, a set of LCs without injected planets, as the goal of this work is to test the ability of different algorithms in recovering bona fide planets rather than determining the \textrm{PLATO}’s planet yield.\\

In the next section, the most successful general-purpose filtering algorithms are briefly described and applied to the samples of \textrm{PLATO} simulated quiet and active stars.

\section{General-purpose algorithms}\label{sec: Hippke algorithms}
\begin{figure}[H]
   \centering

   \includegraphics[width=9cm]{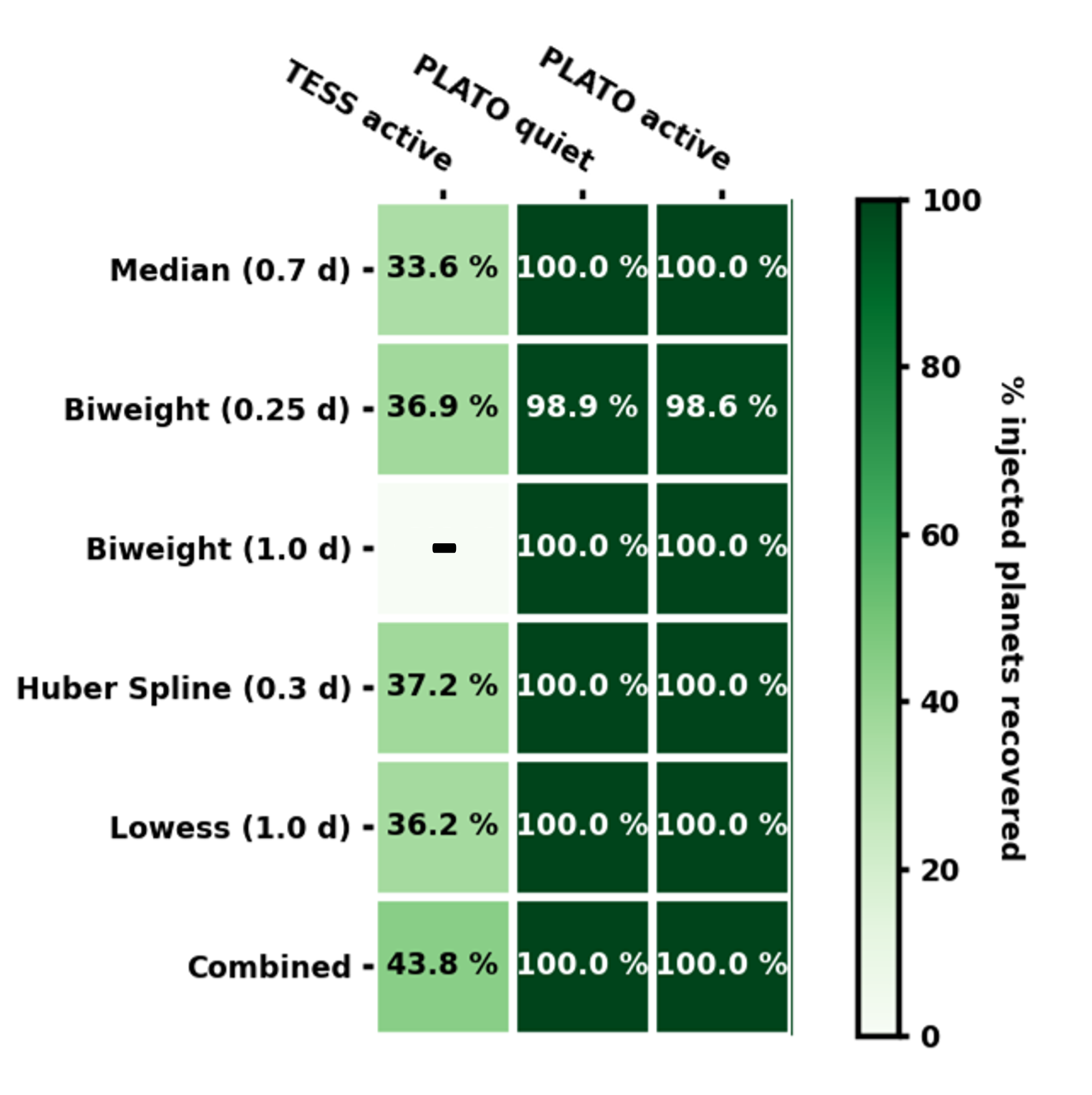}
      \caption{Recovery fraction of injected transit signals similar to those of \citet{Hippke2019} injected into simulated \textrm{PLATO} stellar LCs of quiet (second column) and active (third column) stars, compared with the recovery fraction of similar signals injected into a sample of active stars observed by \textit{TESS} (first column) and analyzed by H19. The biweight with a window of 1.0 d was not tested by H19 on \textit{TESS} active LCs.}
         \label{fig: rec_fraction_Wotan}
   \end{figure}

\begin{figure*}
   \resizebox{\hsize}{!}
            {\includegraphics[width=10cm]{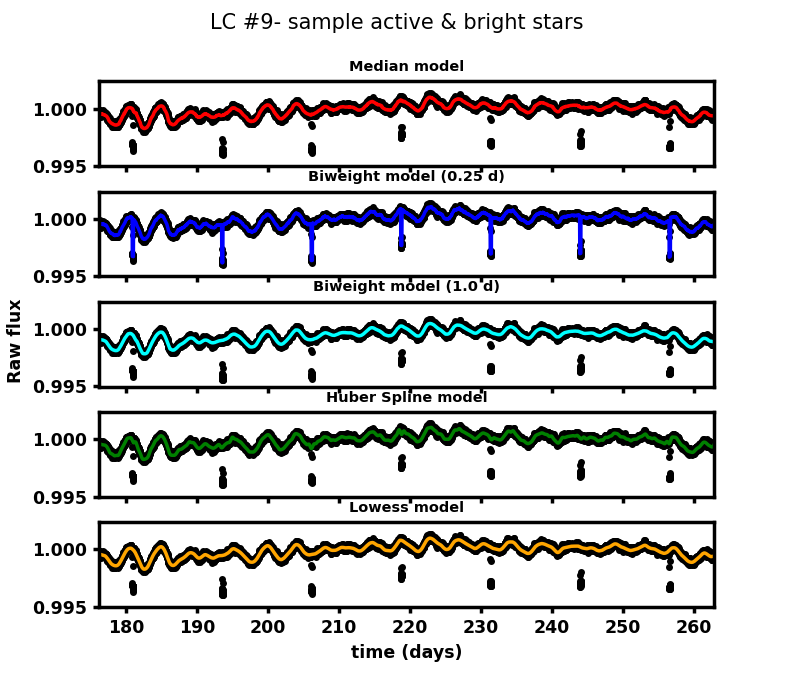}
            \includegraphics[width=10cm]{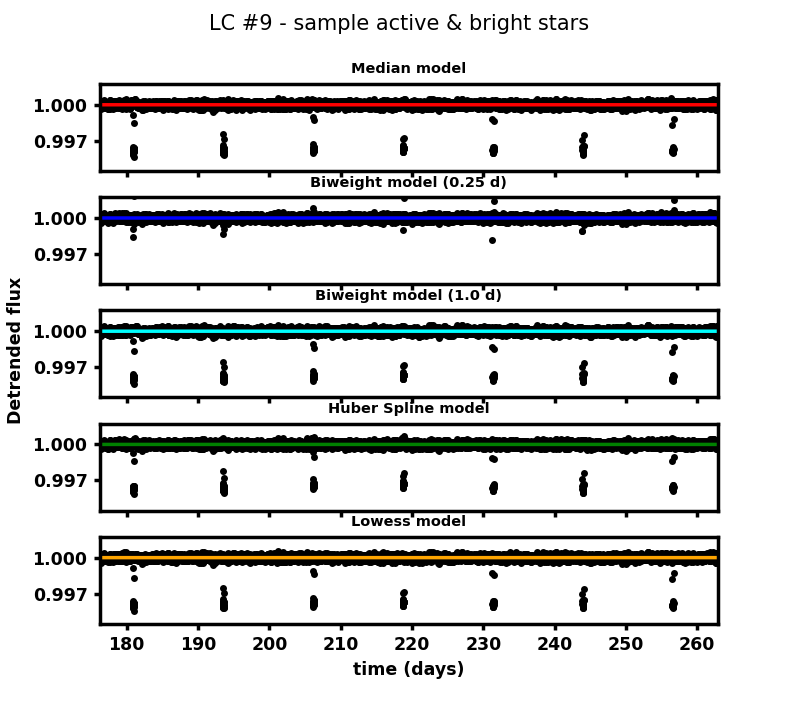}}
      \caption{Raw flux (left panel) and detrended flux (right panel) of the 3rd quarter of LC \#9 of the sample of active stars with an injected transit signal with $P\mathrm{_{orb}=12.60}$ days and $R\mathrm{_p}/R_\star$=0.056. Data-points are shown in black, whereas different colors represent the photometric variability model inferred from algorithm: Median (red), biweight with a window-size of 0.25 days (dark blue) and 1.0 day (light blue), Huber spline (green) and Lowess (orange).}
              
         \label{fig: LC9 model4}
   \end{figure*}

\citet{Hippke2019}, hereafter H19, performed planetary injection-retrieval tests on a sample of 316 highly variable stars observed with \textit{TESS}, obtaining a global detection efficiency lower than 50\% when applying algorithms that were rather successful in detecting small planets around quiet stars. They injected transit signals with the following properties: a planetary radius equal to half the Jupiter radius ($R_\mathrm{p}= 0.5$ $R_\mathrm{J}$), with period randomly drawn from a uniform distribution in the range 1-15 days, and an impact parameter and eccentricity both fixed to zero. We reproduced the same experiment on the samples of both quiet and young active stars of the \textrm{PLATO} simulated LCs, in order to test the performances of the general-purpose filtering algorithms used by H19 (already implemented in the \texttt{W{\=o}tan} python package) on \textrm{PLATO} LCs. In particular, the main aim of this analysis is to investigate how the cadence and the duration of the observations of \textrm{PLATO} can influence the efficiency of the algorithms presented by H19 or if these algorithms are not adequate for the detection of young exoplanets by design. 
Specifically, H19 tested the following algorithms\footnote{We excluded algorithms based on Gaussian Processes from this test as they were first introduced in literature for the filtering of active stars.}: sliding median, Tukey's biweight, Huber spline and Lowess.\\ In the statistical literature, the \textbf{sliding median} (window-size of $w$= 0.7 days) is classified as a "scatterplot smoother" that uses a rectangular (i.e., symmetric and with uniform weights) window kernel of length, $w$, centered on the time, $t(x_i),$ for each data-point, $x_i$.\\ The \textbf{Tukey's biweight} ($w$= 1.0 or 0.25 days for the quiet and active sample, respectively) is a robust location estimator similar to the ordinary least-squares method except that the weights are not constant but depend on the distance from the midpoint, while the loss function\footnote{The loss function is a method to evaluate how well an algorithm models the observed data points. More details can be found in H19.} is given by:
    \begin{equation}
        L(a)=\begin{cases}
        (-(a/c)^2)^2 & \text{if } |a| < c, \\
        0 & \text{otherwise,} \\
        \end{cases}
        \label{eq: biweight}
    \end{equation}
where $a$ are the residuals (i.e., the difference between the observed data and the midpoint) and $c$ is a constant called tuning parameter, usually with a value of $\sim 5$ (default in \texttt{W{\=o}tan}).\\ The \textbf{Huber spline} ($w$= 0.3 days) is a generalization of the maximum likelihood estimation, which is characterized by the following approximated loss function:
    \begin{equation}
        L(a)=c^2(\sqrt{1+(a/c)^2}-1),
        \label{eq: Huber}
    \end{equation}
where $c=1.5$ as the \texttt{W{\=o}tan} default option.\\ Finally, \textbf{Lowess} regression ($w$= 1.0 day) consists in a local polynomial regression method which works by fitting a low-order polynomial to a subset of the data (defined by a certain window-size) at each point along the x-axis using weighted least-squares regression. In this way, more weight is given to the points closer to the data point being estimated.\\
We will refer to these algorithms as general-purpose as they were not specifically developed to deal with prominent stellar activity.
In order to test the efficiency of the aforementioned general-purpose algorithms on \textrm{PLATO} LCs, we performed planetary injection-retrieval tests that were as similar as possible to those done by H19.
For this analysis, we treated each quarter as an independent LC, that is, the test was performed on 800 LCs with 88 days of duration. Planetary transits with the characteristics described above were synthesized with \texttt{batman} and then injected into the single quarters of both the quiet and the active LCs of \textrm{PLATO}. We injected a different planet into each quarter, so that in total, 800 transit signals were analyzed. For the stellar density, $\rho_\star$, and for the limb darkening coefficients, $u_1$ and $u_2$, we adopted the values reported in the tables describing the simulated LCs, provided by the LSWG (shown in Appendix B). 
Afterward, the LCs with injected transits were filtered with the \texttt{W{\=o}tan} algorithms with the same window-sizes adopted by H19. Finally, we analyzed the filtered fluxes by searching for planetary transits with TLS.\\
The general-purpose algorithms achieved unexpectedly excellent results on both samples of the \textrm{PLATO} simulated LCs, when compared to the results presented in H19.
The percentage of recovered transits with TLS after the filtering with these algorithms is reported in Fig. \ref{fig: rec_fraction_Wotan}, where each row represents the results of a different algorithm, and each column refers to a different sample of LCs. Specifically, in the first column the results obtained by H19 on \textit{TESS} active stars are reported, whereas in the second and third columns, the results from the \textrm{PLATO} quiet and active sample are shown, respectively. In the last row, the recovery percentage obtained combining all methods is also shown. As can be seen from this figure, the recovery efficiency of the tested general-purpose algorithms on both samples of \textrm{PLATO} stellar LCs is extremely high, recovering 100\% of the injected signals in almost all the cases. The only exception is the biweight algorithm with a window-size of 0.25 days, which recovers 98.88\% and 98.62\% of the transit signals in the quiet and active sample, respectively.
In a reanalysis of the LCs of both samples with the biweight algorithm and a window-size of 1 day -- a good approximation of the best value proposed by H19 (three times the duration of the transit) for orbital periods up to 15 days -- the TLS recovers the totality of the injected transit signals.\vspace{-15pt}
\vspace{-10pt}
An example of a planet missed by TLS after filtering the LC with the biweight with $w=0.25$ days is given in Fig. \ref{fig: LC9 model4}, where the raw and the detrended flux of quarter \#3 of LC \#9 of the active sample are shown. In this quarter, a planet with an orbital period ($P\mathrm{_{orb}}$) of 12.60 days and a scaled planetary radius ($R\mathrm{_p}/R_\star$) of about 0.056 is injected. In each panel of this figure, the stellar variability model obtained by means of different algorithms is highlighted in different colors: red for the median, green for the Huber spline, orange for the Lowess, blue for the biweight (specifically, dark blue for a window-size of $w=0.25$ d and light blue for $w=1.0$ day). It can be easily observed that the biweight with the shortest window fails in correctly filtering the LC, so that in the corresponding corrected flux, almost all the transits disappear.  
Nevertheless, by increasing the window-size up to 1.0 day, transit features are perfectly recognizable after the filtering and the transit is correctly detected.\\

\begin{figure}[H] 
\centering
\includegraphics[width=9cm]{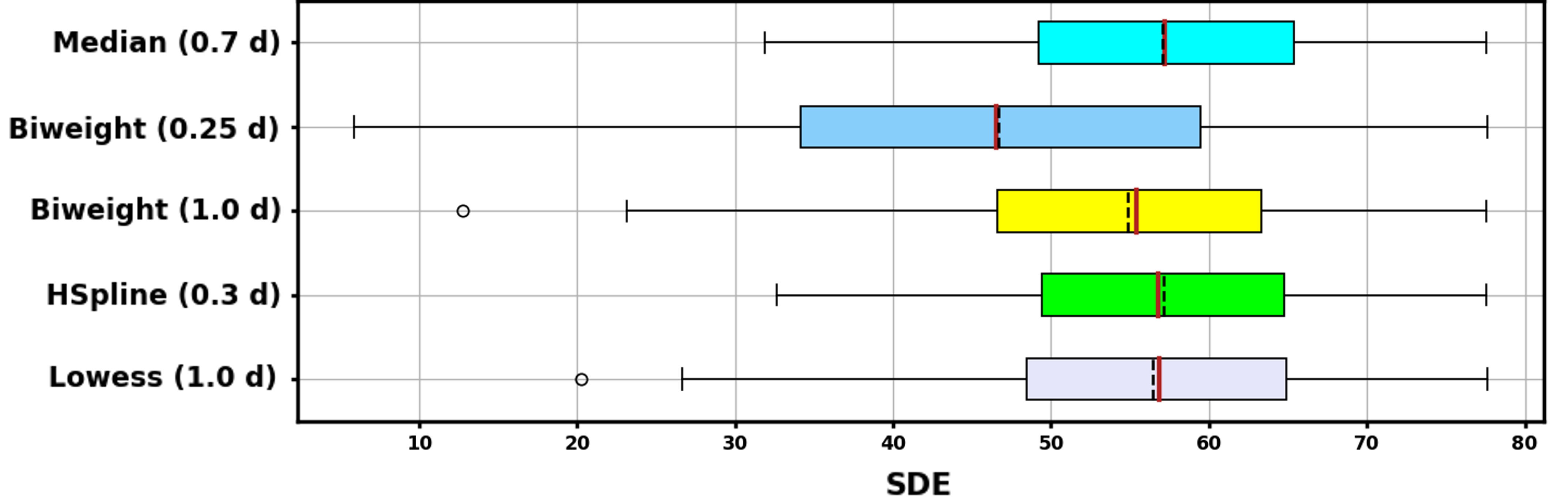}\hfil \centering
\caption{Boxplot of the signal detection efficiency for the general-purpose algorithms from \texttt{W{\=o}tan} tested on the active sample with injected transit signals with $R\mathrm{_p} \sim 0.5 R\mathrm{_J}$ and $P\mathrm{_{orb}} \in [1-15]$ days, moving on circular orbits. Boxes cover the lower to upper quartiles; whiskers show the 10 and 90 percentiles. Dashed black lines indicate the mean SDE, whereas red lines the median. The black circles in the biweight (1.0 d) and the lowess (1.0 d) cases represent data that extend beyond the whiskers.}\label{fig: SDE_activeW}
\centering
\end{figure}

Considering the SDE of the recovered transits computed by TLS in the active sample, in the boxplot in Fig. \ref{fig: SDE_activeW}, the main properties of the SDE distributions are summarized. It can be immediately visualized that the SDE for most of the algorithms ranges between $\sim 30$ and $\sim 76$, whereas the biweight with the shortest window-size reaches much lower values. 
Similar results are obtained from the analysis of the quiet sample of \textrm{PLATO} simulated LCs.

Although we cannot exclude that the level of activity in the \textrm{PLATO} LCs is lower than those observed in the \textit{TESS} dataset analyzed by H19, our analysis shows that the low efficiency general-purpose algorithms for the filtering of active stars derived by H19 are likely driven by the characteristics of the LCs, that is, the longer time span and the noise properties. For this reason, we decided to include the two\footnote{We have to select two algorithms due to the limited computing resources and for the sake of readability, although they all have similar performances.} best-performing algorithms in our comparison and in the analysis presented in H19.
Overall, the biweight with $w=1.0$ d and the Huber spline with $w=0.3$ d appear to be the algorithms with the highest values of SDE and the lowest standard deviation, with regard to the quiet and active sample, respectively. Therefore, these two algorithms are tested also on the other injection-retrieval experiments performed, considering also smaller planets, as explained in the next sections.

\section{Custom-built algorithms}\label{sec: new algorithms}
As previously mentioned in Sect. \ref{sec: intro}, four independent research groups have recently developed algorithms specifically for the filtering of activity in the LCs of young stars observed by \textit{K2} and \textit{TESS}. In this work, they were implemented in Python and applied over the two samples of \textrm{PLATO} simulated LCs. A brief description of the selected algorithms is given below.\\ 
\textbf{N\&L}, developed by \citet{Rizzuto2017}, is a two-step pipeline, called \texttt{Notch} and \texttt{LOCoR}, respectively. \texttt{Notch} works by fitting two models over a moving data window: a simple quadratic polynomial and a polynomial with a box-shape transit notch, in order to identify the possible presence of a transit. Then, the model with the lowest Bayesian Information Criterion (BIC) is adopted.
The selection of the window-size employed for the filtering of \textrm{PLATO} LCs is summarized in Eq. \ref{eq: window_size} and it depends on the stellar rotation period $P\mathrm{_{rot}}$:
\begin{equation}
    {\rm window-size} (\rm days)= 
    \begin{cases}
    2.0 & \text{if } P_\mathrm{rot}>13 \text{ days,} \\
    1.0 & \text{if } 2 < P_\mathrm{rot} \leq 13 \text{ days,}\\
    0.5 & \text{if } P_\mathrm{rot}\leq 2 \text{ days.}
    \end{cases}
    \label{eq: window_size}
\end{equation} 
For $P\mathrm{_{rot} \leq 2}$ days, \texttt{Notch} is unable to completely remove rotational systematics, therefore the \texttt{LOCoR} algorithm should also be applied. The latter is based on the \texttt{LOCI} algorithm of \citet{LOCI} and it consists of modeling each individual rotation as a linear combination of other rotations of the same star in the dataset.\\
\textbf{YSD}, developed by \citet{YSD}, is a Python pipeline which employs Lowess smoothing regression to model stellar variability, using the standard tricube as a weighting function:
\begin{equation}
        w(x)=(1-|d|^3)^3
        \label{eq: tricube}
\end{equation}
where the weight $w$ at each data point, $x,$ is given by the distance, $d,$ from the point on the curve being fitted, scaled to lie in the range [0,1].\\ 
Therefore, the detrending pipeline is carried out by cutting the outliers\footnote{The other algorithms do not include an outliers removal.} (peaks and troughs) first, and then by estimating the variability trend using the Lowess function implemented in the Python module \texttt{statsmodel} (\citealt{seabold2010statsmodels}) with frac=0.012 when considering one quarter only, meaning that a fraction of 1.2\% of the data has been used to estimate each value of the variability model. When filtering the entire \textrm{PLATO} LCs, the frac parameter is set to 0.0015 instead, because it results in better performances.\\
The outliers are identified with the function \texttt{find\_peaks} in the \texttt{scipy.signal} (\citealt{SciPyref}) package\footnote{ https://docs.scipy.org/doc/scipy/reference/signal.html}, using a prominence of 0.001 and a width of 20 data points. These parameters should be adjusted for more complex LCs. For more information about the outlier cutting see Sect. 2.5.2 of \citet{YSD}.\\
\textbf{K2SC}, developed by \citet{Aigrain2016}, is one of the filtering algorithms that makes use of Gaussian processes (GPs: \citealt{RasmussenW06}) to model both instrumental systematics and astrophysical variability. Overall, GPs are non-parametric methods which describe a dataset of $N$ data points by evaluating correlations between them through a kernel or covariance function. The latter describes how each point is related to all the other points, and this relation is expressed through a $N \times N$ matrix, called a covariance matrix.
To speed up computation time, we employed the covariance functions from the \texttt{celerite2}\footnote{https://github.com/exoplanet-dev/celerite2} (\citealt{celerite2}) package in our implementation of Gaussian processes for the \textrm{PLATO} LCs.
    
Activity is modeled as a mix of two stochastically driven and damped harmonic oscillators (SHO), describing two modes in Fourier Space at the stellar rotation period, $P\mathrm{_{rot}}$, and its first harmonic, respectively, and implemented in the code as the \textit{RotationTerm} kernel.
It was shown that this kernel is able to reproduce a large range of stellar variability in time-series dataset, from stellar rotation to pulsations (e.g., \citealt{Haywood}, \citealt{Rajpaul2015}, \citealt{Uttley}). Besides the \textit{RotationTerm} ($K_\mathrm{rot}$), describing the time-dependent variability of the light curve, the \textit{JitterTerm} ($K_\mathrm{wn}$) implemented in \texttt{celerite2} is added to model the non-periodic variability or white noise, This second term is given by $K_\mathrm{wn,ij}=\sigma^2 \delta_{ij}$, where $\delta_{ij}$ is the Kronecker delta function. Therefore, the final covariance function, defining each element of the covariance matrix with indices ($i,j$), is given by: $K_\mathrm{ij}=K_\mathrm{rot,ij}+K_\mathrm{wn,ij}$.  
The filtering has been done by means of the \texttt{exoplanet}\footnote{https://github.com/exoplanet-dev/exoplanet} package (\citealt{exoplanet:exoplanet}), fitting the data and imposing priors through the \texttt{PyMC3}\footnote{https://docs.pymc.io/} (\citealt{exoplanet:pymc3}) statistical distributions, following \citet{Aigrain2016} as much as possible, even if the adopted kernels were quite different from the original pipeline of K2SC. For this reason, from now on, we will refer to this filtering algorithm as \textbf{GPs} and not K2SC. Despite the general name adopted in this work (GPs), it is important to highlight that Gaussian processes can be tested with many different kernels. In this work, we investigate only the specific kernel described above and a different kernel would likely return different results.\\
We adopted the following priors over the hyperparameters of the covariance function, with distributions defined in \texttt{PyMC3}:
 1) the primary period of variability, fixed to the rotation period $P_\mathrm{rot}$ given in tables describing the simulated LCs, provided by the LSWG; 
2) the standard deviation of the process $\sigma$, equivalent to the amplitude of the variability, with uniform prior in logarithmic space between $10^{-7}$ and $10^{0}$ in unit of normalized flux (i.e., flux divided by its median); 
3) the quality factor of the secondary oscillation minus one half \textit{Q$_0$}, with normal prior with mean 0.0 and standard deviation 2.0  in logarithmic space;
4) the difference between the quality factors of the first and second modes \textit{dQ}, with normal prior with mean 0.0 and standard deviation 2.0 in logarithmic space;
5) the fractional amplitude of the secondary mode compared to the primary \textit{f}, with uniform prior between 0.1 and 1.0.\\   
\textbf{VARLET}, developed by \citet{VarletPhalet}, is a wavelet-based filtering pipeline divided into three main steps. Firstly, the raw flux is decomposed in wavelets at multiple levels through the discrete wavelet transform (DWT). Then, a thresholding function is applied to the details coefficients obtained from the previous multilevel decomposition. Specifically, a soft-thresholding procedure has been adopted (\citealt{Donoho1994}), meaning that the corrected version ($Y$) of the original detail coefficients ($X$) is given by: $Y=sign(X) \cdot f(|X|-thr)$, where $thr$ is the chosen threshold and $f(|X|-thr)$ is defined by the following equation:
    \begin{equation}
        f(|X|-thr)= 
        \begin{cases}
        |X|-thr & \text{if } |X|\geq thr, \\
        0 & \text{otherwise,} \\
        \end{cases}
        \label{eq: soft_thresholding}
    \end{equation}
    where the threshold is computed through the penalized criterion provided by \citet{birge2007}. 
    Finally, the denoized light curve is reconstructed through the inverse discrete wavelet transform (IDWT) using the corrected coefficients. Therefore, the reconstructed LC, which should represent stellar variability, is subtracted from the raw flux, thus obtaining a residual LC containing only the transit signals and the white noise.\\
    Originally written in MATLAB, the pipeline has been re-implemented in Python by adopting the \texttt{PyWavelets}\footnote{https://github.com/PyWavelets/pywt} package (\citealt{Lee2019}). In the original pipeline, the VARLET filter is usually used with a standard BLS which does not take the transit depth or the shape (other than boxlike) into account. In this work, the search for transit signals in the filtered LC is instead carried out by means of the TLS.\\
All these algorithms have been tested together with the biweight and Huber spline algorithms described in Sect. \ref{sec: Hippke algorithms}.

\section{Injection-retrieval tests}\label{sec: tests}
\begin{figure*}
   \resizebox{\hsize}{!}
{\includegraphics[width=.5\textwidth]{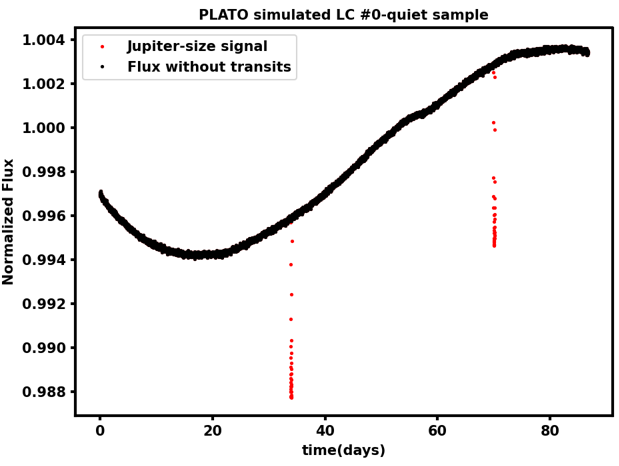}\hfil 
\includegraphics[width=.5\textwidth]{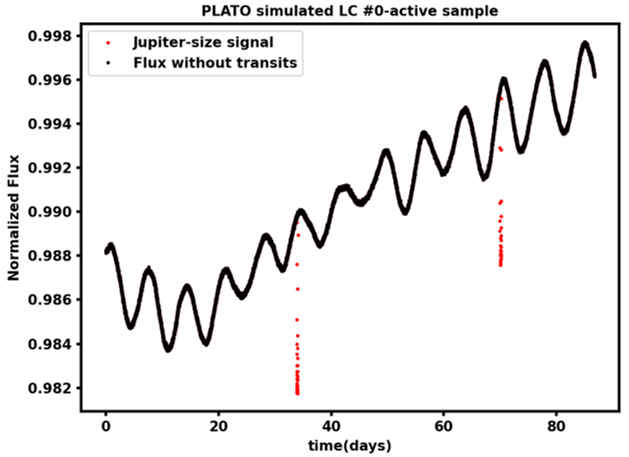}\hfil}
\vspace{-20pt}
\caption{First quarter of the \textrm{PLATO} simulated LC \#0 of the quiet (left) and active (right) sample, with an injected Jupiter-size planet. In black, we show the normalized flux without transits and in red, the transit model generated with \texttt{batman}.}\label{fig: PLATOLC_example_NJ}
\centering
\end{figure*}

\begin{figure*}
   \resizebox{\hsize}{!}
{\includegraphics[width=.5\textwidth]{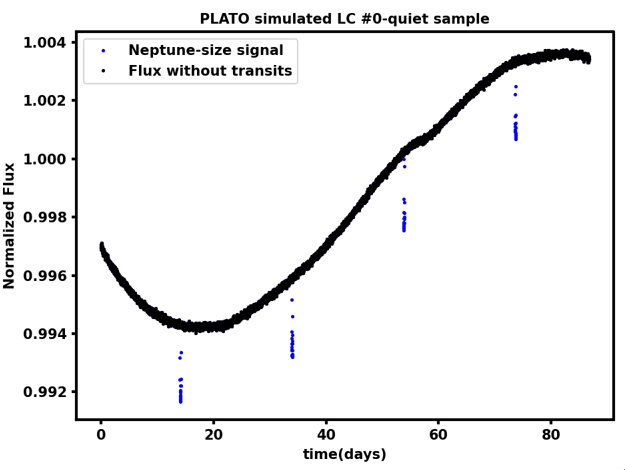}\hfil 
\includegraphics[width=.5\textwidth]{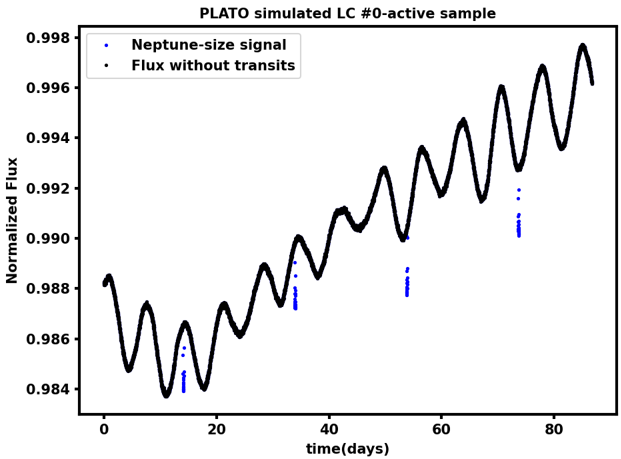}\hfil}
\vspace{-20pt}
\caption{First quarter of the \textrm{PLATO} simulated LC \#0 of the quiet (left) and active (right) sample, with an injected Neptune-size planet. In black, we show the normalized flux without transits and in blue, the transit model generated with \texttt{batman}.}\label{fig: PLATOLC_example_MNSE}
\centering
\end{figure*}

\begin{figure*}
   \resizebox{\hsize}{!}
{\includegraphics[width=.5\textwidth]{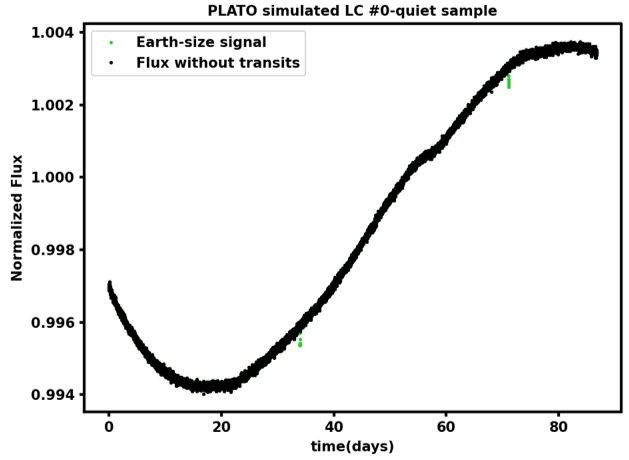}\hfil 
\includegraphics[width=.5\textwidth]{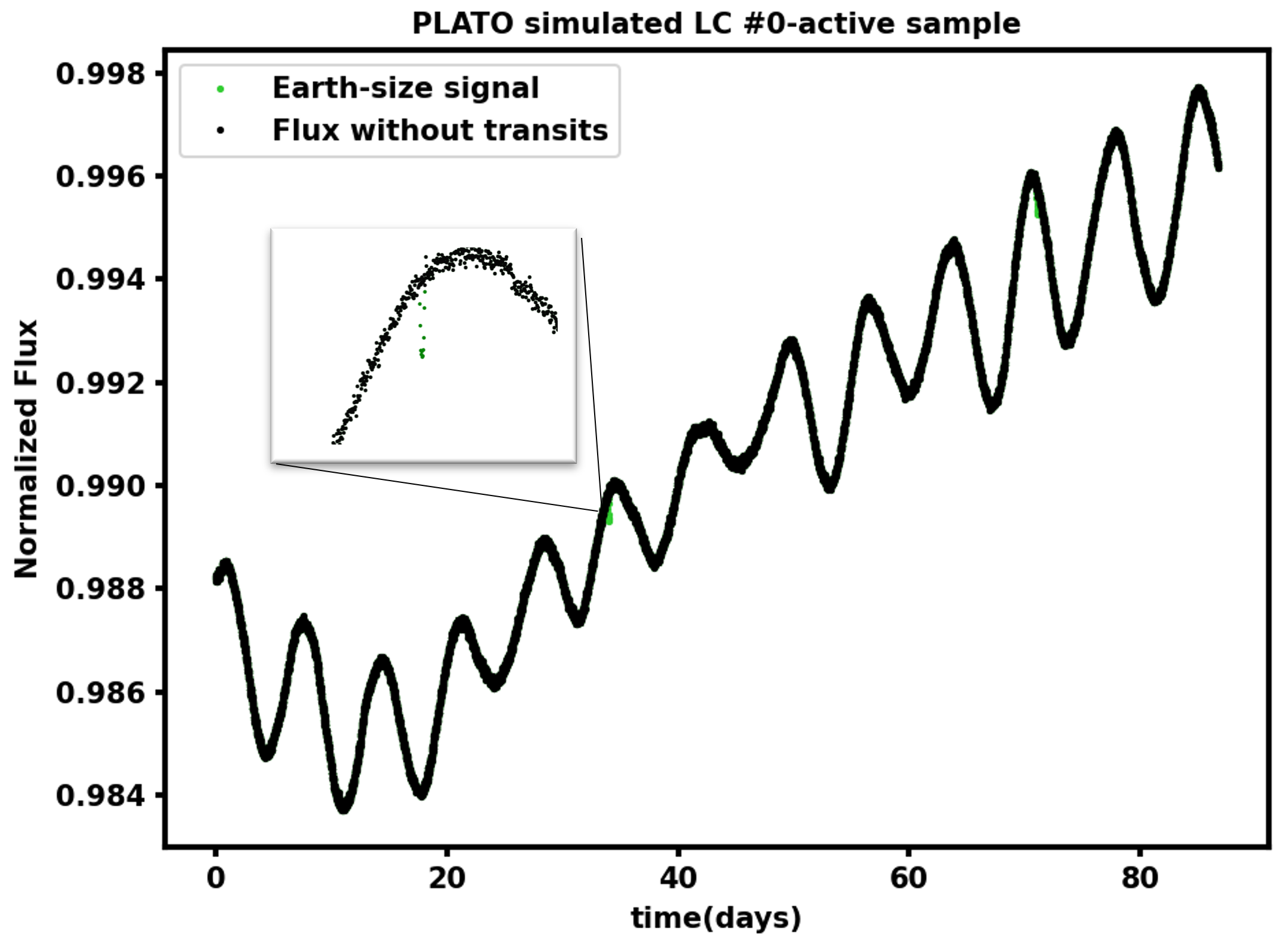}\hfil}
\vspace{-20pt}
\caption{First quarter of the \textrm{PLATO} simulated LC \#0 of the quiet (left) and active (right) sample, with an injected Earth-size planet. In black, we show the normalized flux without transits and in green, the transit model generated with \texttt{batman}. In the active case, the transits are hard to distinguish, therefore a zoom-in on the first transit is shown. The second transit occurs at about 70 days.}\label{fig: PLATOLC_example_E}
\centering
\end{figure*}

In order to test the transit detection efficiency of the filtering algorithms described above, a series of planetary transit signals were synthesized and injected into the \textrm{PLATO} simulated LCs before filtering.\\
Transits representing planets with different scaled planetary radii ($R\mathrm{_p}/R_\star$) in the range 0.01-0.1, orbiting their host star in circular orbits (i.e., zero eccentricity), with an orbital period ($P\mathrm{_{orb}}$) between 0.75-40 days and an impact parameter ($b$) drawn from a uniform distribution in the range [0-1], have been synthesized with the \texttt{batman} package and then injected into single quarters of both the quiet and active sample. 
We sampled the scaled planetary radius $R\mathrm{_p}/R_\star$ from uniform distributions in the following ranges ($\mathcal{U}(a,b)$):
1) Jupiter-size: $R\mathrm{_p}/R_\star$ from $\mathcal{U}(0.05,0.1)$; 2) Neptune-size: $R\mathrm{_p}/R_\star$ from $\mathcal{U}(0.03,0.05)$; 3) Earth-size: $R\mathrm{_p}/R_\star$ from $\mathcal{U}(0.01,0.03)$.\\
Each planetary type (Jupiter-, Neptune-, and Earth-size) was analyzed independently from the other, that is: each LC was injected and analyzed three times (once per each planetary type), without mixing the results.
Examples of the first quarter of LC \#0 of both the quiet and active samples of \textrm{PLATO} simulated LCs, with injected transits for a Jupiter-, Neptune-, and Earth-like signal, are shown in Fig. \ref{fig: PLATOLC_example_NJ}, \ref{fig: PLATOLC_example_MNSE}, and \ref{fig: PLATOLC_example_E}, respectively.\\
As previously done, we retrieved the limb darkening coefficients and the stellar density ($\rho_\star$), which is in the range 0.8-3.0 $\mathrm{g/cm^3}$, taken from tables provided by the LSWG.
The scaled semi-major axis ($a/R_\star$) were derived from the orbital period and the stellar density through Kepler's third law.\vspace{-20pt} 
\vspace{-15pt}
\vspace{-20pt}
\vspace{-20pt}

\clearpage
\subsection{Individual quarters}\label{sec: individual_quarters}
\begin{figure}[H]
   \centering
   \includegraphics[width=9cm]{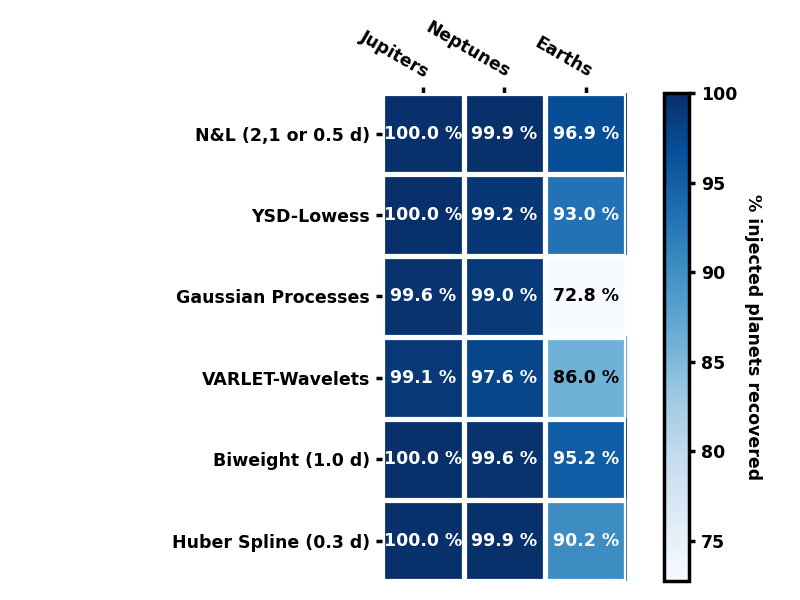}
      \caption{Recovery fraction of injected transit signals in simulated \textrm{PLATO} stellar LCs of the sample of quiet stars.}
         \label{fig: rec_fraction_quiet1}
   \end{figure}

\begin{figure}[H]
   \centering
   \includegraphics[width=9cm]{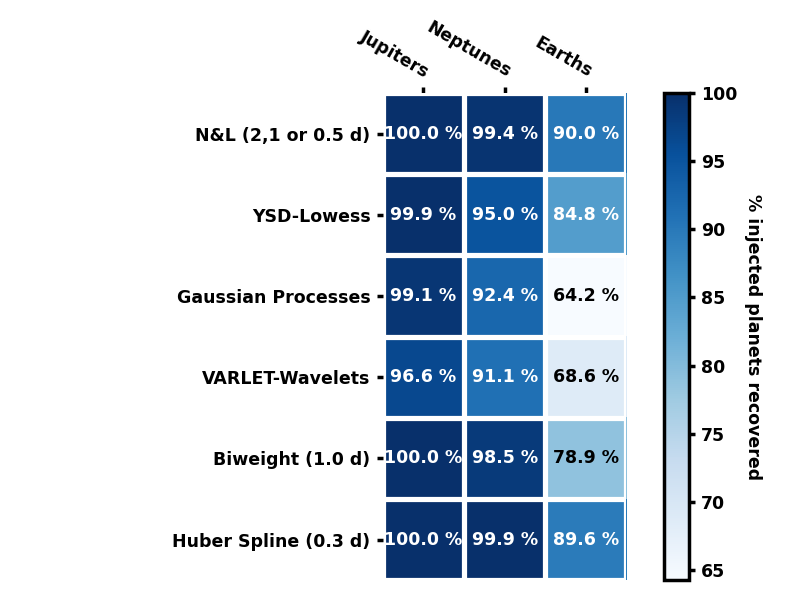}
      \caption{Recovery fraction of injected transit signals in simulated \textrm{PLATO} stellar LCs of a sample of active stars.}
         \label{fig: rec_fraction_active1}
   \end{figure}

We performed the first transit injection-retrieval test considering each quarter of a simulated LC independently from the others, for a total of 800 tests for each planetary type (Jupiter-, Neptune-, and Earth-size type) as each of the 100 LCs comprises eight quarters. Although the eight quarters from the same LC are sharing the same stellar parameters, the different instrumental systematics and the quasi-periodic and non-stationary nature of stellar activity characterize each quarter in a different way, thus justifying our choice in the framework of testing the efficiency of algorithms. A given transit model is then injected in both the quiet and active samples, while keeping the samples separated. It follows that each efficiency rate is computed over 800 tests.\\
Each LC is then filtered using the algorithms described in Sect. \ref{sec: new algorithms} and the two general-purpose algorithms (i.e., the biweight and the Huber spline).
The percentage of recovered transits with TLS after the filtering with these algorithms is reported in Figs. \ref{fig: rec_fraction_quiet1} and \ref{fig: rec_fraction_active1}, for the quiet and active sample, respectively. Each row represents the results of a different algorithm, whereas each column refers to a different planetary type, as in the legend. 
\vspace{-10pt}

   \vspace{-10pt}

First of all, it can be noticed that in the quiet sample all the algorithms perform quite well for Jupiter- and Neptune-sized planets, as all filtering algorithms lead to transit recovery rates with TLS higher than 99.0\%, except for VARLET which achieves 97.6 \% for the Neptune class. For Earth-sized planets (i.e., $R\mathrm{_p}/R_\star \leq 0.03$), all algorithms achieve retrieval rates higher than 90\% except for VARLET (86\%) and GPs (72.8\%).\\ 
With regard to the active sample, a slightly decrease in the number of recovered transits has been observed, as expected, due to the much higher photometric variability of these LCs. Nevertheless, among the custom-built algorithms,  N\&L results to be the best-performing algorithm for all planetary types in both samples, recovering 96.87\% and 90\% of small planets in quiet and active stars, respectively. Among the general-purpose algorithms, the Biweight and the Huber spline have performances extremely close to those of N\&L in the quiet sample and the active sample respectively. 
Among the algorithms with a lower detection efficiency, we find the VARLET code for what regards the larger planets, that is, Jupiters and Neptunes, while the GPs slightly underperform with respect to the other methods for what regards the shallow transits generated by small planets.

\begin{figure*}
   \resizebox{\hsize}{!}
            {\includegraphics[width=10cm]{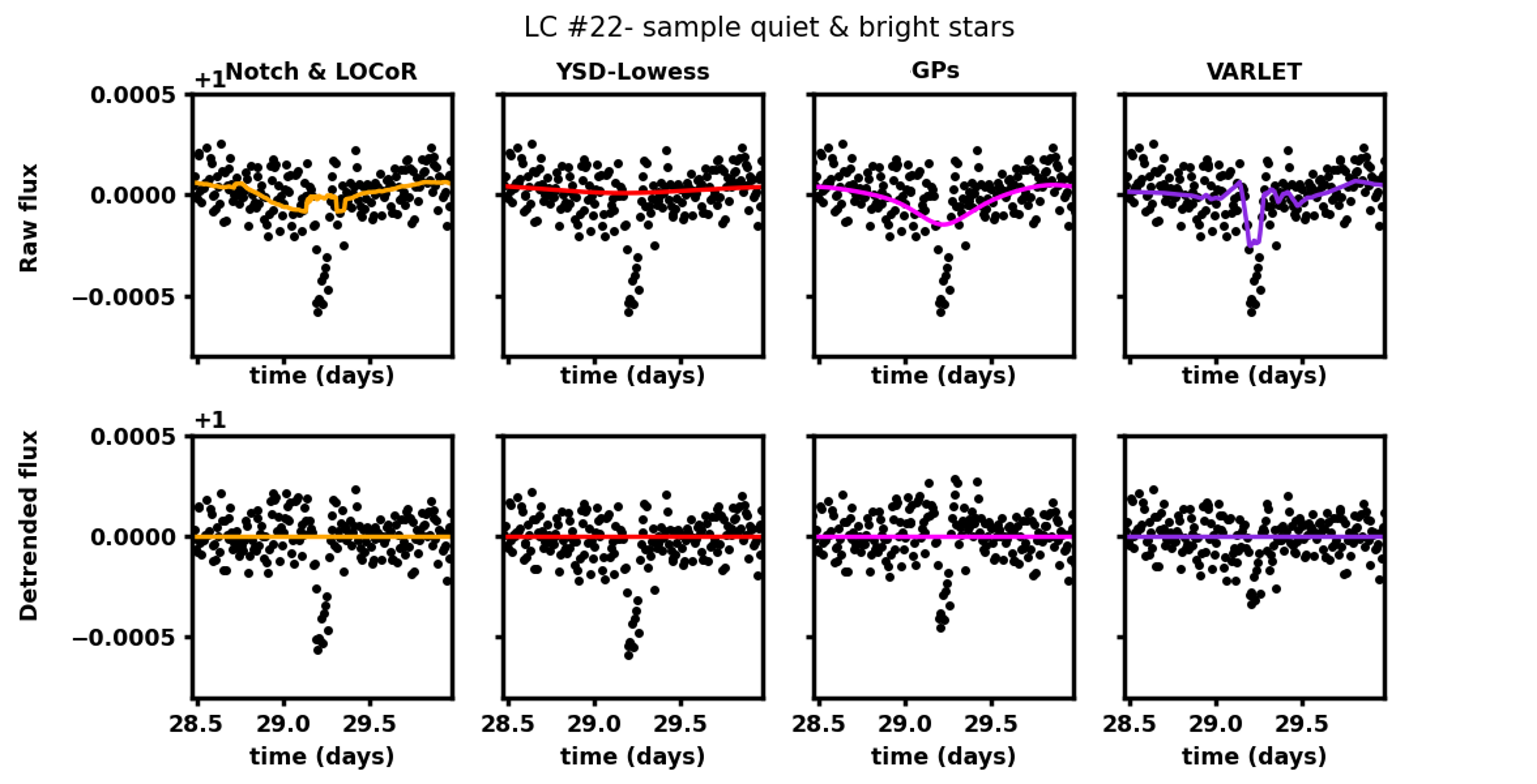}}
      \caption{Raw flux (top) and detrended flux (bottom) of the first quarter of LC \#22 of the quiet sample, zoomed in on the transit points of the injected Earth-size planet. Data points are shown in black, whereas different colors represent the photometric variability model inferred from different algorithms: N\&L (orange), YSD (red), GPs (fuchsia), and VARLET (violet).}
              
         \label{fig: LC22 zoom}
   \end{figure*}

H19 observed that GPs tend to absorb the transit signal by modeling it as part of the photometric variability of the light curve, thus reducing the transit depth as well as the sensitivity in the subsequent transit search in the filtered flux. This behavior was noticed also in the tests on \textrm{PLATO} LCs, not only with the GPs, but also with VARLET.
An example is given in Fig. \ref{fig: LC22 zoom}, where the Earth-like transit points in LC \#22 are shown, with overplotted variability models inferred from different algorithms. It can be easily noticed that the filtering models determined by either the GPs (third top panel on the right, fuchsia line) or the VARLET model (last top panel on the right, violet line) exhibit a dip in correspondence of the transit, which is then shallower in the detrended flux (bottom panels), thus reducing the sensitivity and the SDE in the subsequent transit search. In the case of VARLET, this behavior is observed also for giant planets with large transit depths.\\ In general, we expect higher values of the SDE in the detection of planets with larger size (i.e., deeper transits) or shorter orbital periods (i.e., with  a higher number of transits). This behavior was observed for all the filtering algorithms -- except for VARLET, which generates a SDE distribution with similar low values regardless of the planetary size, as a consequence of systematically reducing the transit depths of all transits.

\begin{figure}[H]
   \centering
   \includegraphics[width=9cm]{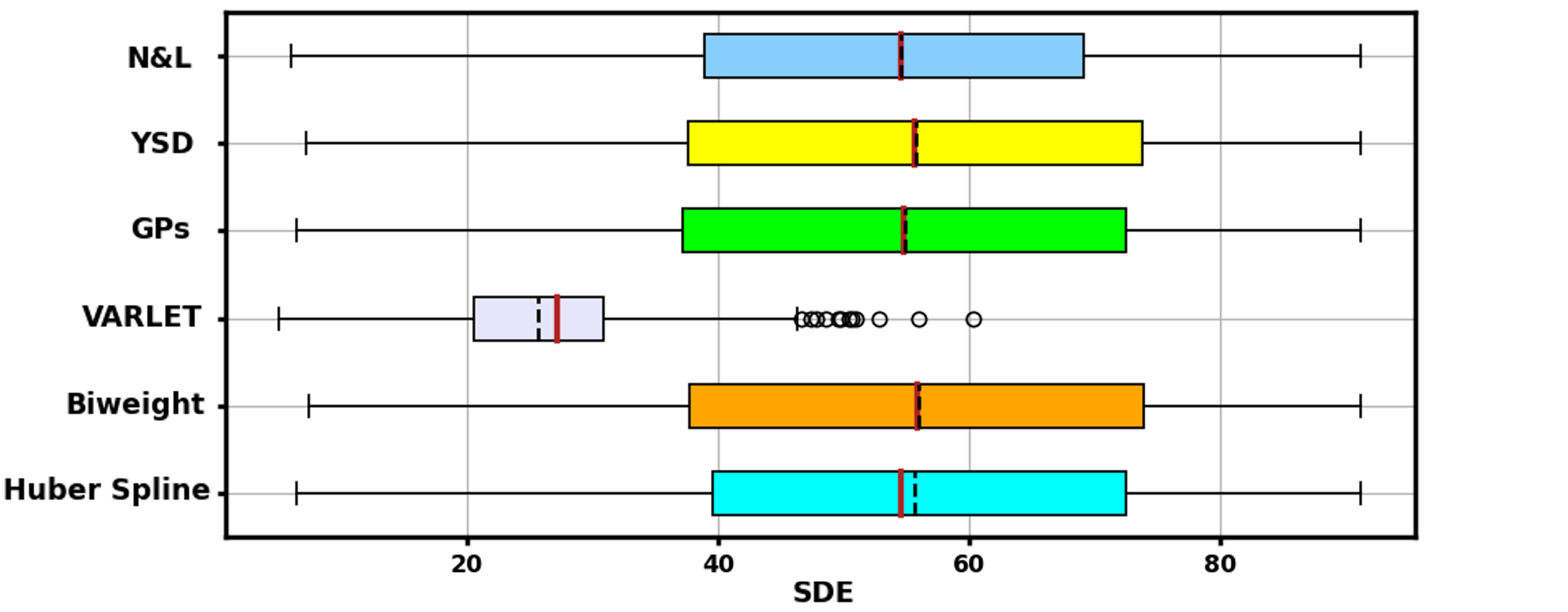}
      \caption{Boxplot of the SDE for the algorithms listed on the y-axis, tested on the quiet sample with injected transit signals as described in Sect. \ref{sec: individual_quarters}. Boxes cover the lower to upper quartiles; whiskers show the 10 and 90 percentiles. Dashed black lines indicate the mean SDE, whereas red lines the median. The black circles in the VARLET case represent data that extend beyond the whiskers.}\label{fig: SDE_quiet1}
   \end{figure}

\begin{figure}[H]
   \centering
   \includegraphics[width=9cm]{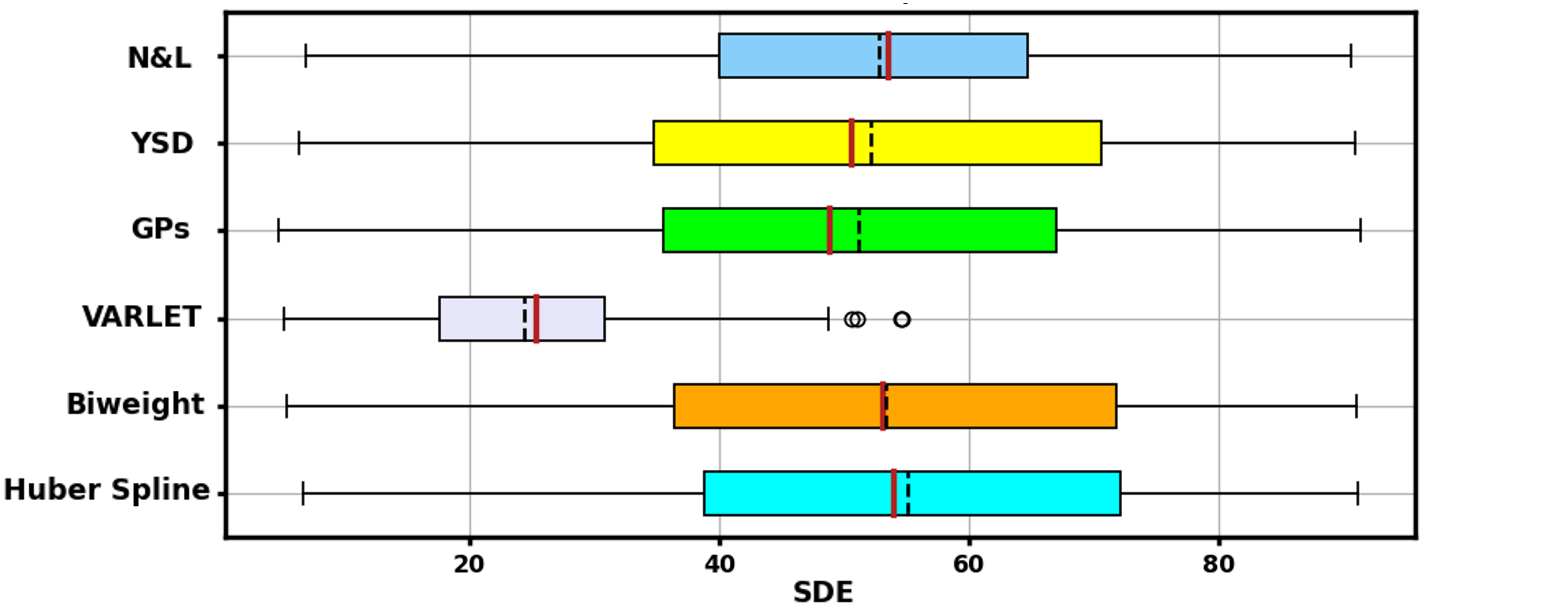}
    \caption{Boxplot of the SDE for the algorithms listed on the y-axis, tested on the active sample with injected transit signals as described in Sect. \ref{sec: individual_quarters}. Boxes cover the lower to upper quartiles; whiskers show the 10 and 90 percentiles. Dashed black lines indicate the mean SDE, whereas red lines the median. The black circles in the VARLET case represent data that extend beyond the whiskers.}
      \label{fig: SDE_active1}
   \end{figure}

In particular, VARLET performs weakly in terms of SDE, achieving a maximum value of $\sim$54, whereas all the other algorithms show similar performances, with SDE values up to 90 for the large planets. Similar results are obtained from the filtering done by the biweight and the Huber spline. Figures \ref{fig: SDE_quiet1} and \ref{fig: SDE_active1} show a performance summary of the analyzed methods on the quiet and active sample, respectively. With regard to the SDE of the general-purpose algorithms, it can be noticed that the biweight performs slightly better than the Huber spline in the case of quiet stars, whereas the situation is reversed in the case of active stars.\\ 
Finally, we note that the SDE of the detected transits on the filtered LCs of the active stars are similar to those of the quiet sample. This is an encouraging result, since it suggests that the activity filtering algorithms perform equally well for the two datasets, even for stars that are very different from the photometric point of view. This in turn suggests that \textrm{PLATO} will be able to detect a large number of planets with very different properties both in terms of size and periods, also around young or photometrically active stars. 
   \vspace{-10pt}

\subsection{Caveats in transit detection}
\begin{figure}[H]
\centering
\includegraphics[width=9cm]{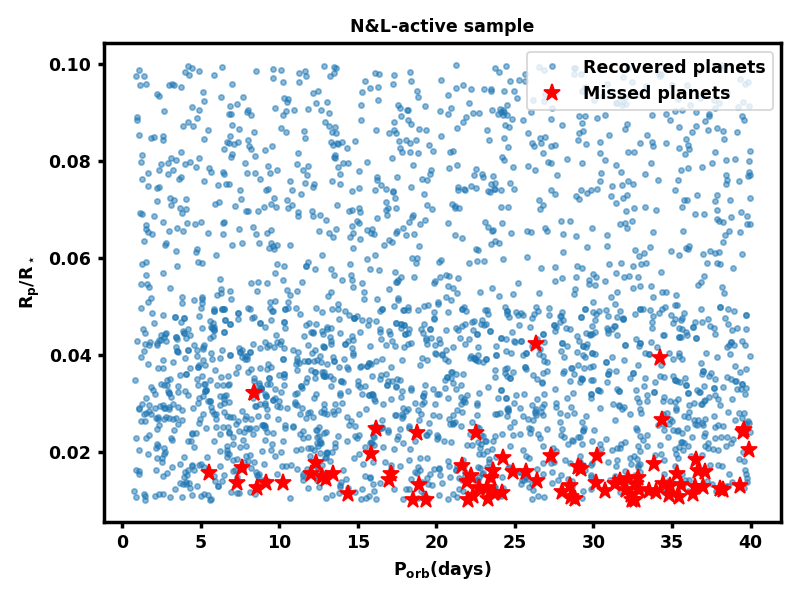}\hfil
\includegraphics[width=9cm]{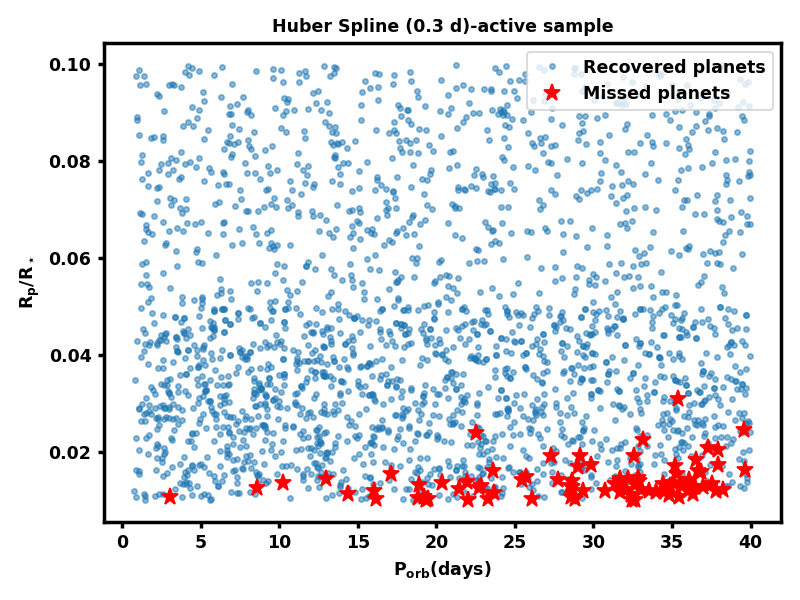}\hfil 
\centering
\caption{Injection-recovery tests on the sample of active stars. Injected scaled planetary radius as a function of the injected orbital period: blue dots are the recovered planets, whereas red stars are the missed planets after filtering the LCs with N\&L (top) and Huber spline (bottom).}\label{fig: Rp_Rs vs Porb active}
\centering
\end{figure}

An in-depth analysis of the parameter space of the recovered planets shows that Jupiter and Neptune-size planets are mostly recovered regardless of their orbital period. On the other hand, Earths and super-Earths are harder to identify, especially at long orbital periods, since the number of transit signals decreases with increasing $P\mathrm{_{orb}}$. This is to be expected since Earth-like planets produce shallow transit signals in the LCs, which can be overcome by the variability amplitude. An example of this trend is shown in Fig. \ref{fig: Rp_Rs vs Porb active}, where the scaled planetary radius of injected planets is plotted as a function of the injected orbital period. Blue dots represent planets recovered by the TLS in the LCs of the active sample filtered by N\&L (top) or Huber spline (bottom).

\begin{figure}
\centering
\includegraphics[width=9cm]{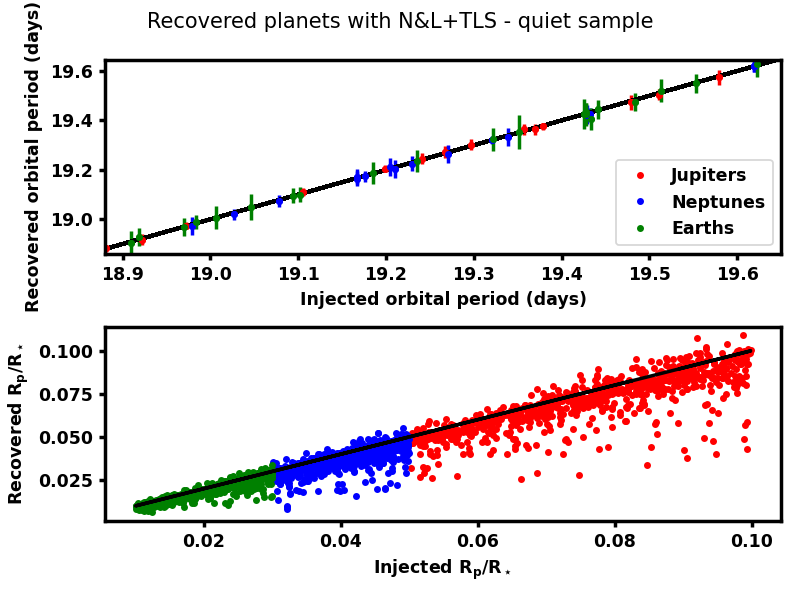}\hfil
\centering
\caption{Results of the planetary injection-recovery tests after filtering the LCs of the quiet sample with N\&L. Upper panel: Recovered orbital period ($P\mathrm{_{orb}}$) as a function of the corresponding injected $P\mathrm{_{orb}}$. Lower panel: Recovered scaled planetary radius ($R\mathrm{_p}/R_\star$) as a function of the corresponding injected $R\mathrm{_p}/R_\star$, for the three planetary signals, colored as in the legend. Black solid line marks points where all the data should lay if the recovered values were exactly the same as the injected ones.}\label{fig: inj_vs_detNL}
\centering
\end{figure}

Furthermore, considering the orbital and planetary parameters of the recovered planets measured by TLS, it is found that filtering algorithms allow to correctly detect the mid-transit time and to measure a very precise estimate of the planetary orbital period with small uncertainties. However, this is not valid anymore when considering the scaled planetary radius $R\mathrm{_p}/R_\star$, which is often underestimated,  especially for large planets (i.e., $R\mathrm{_p}/R_\star \geq 0.05$). As an example, in Fig. \ref{fig: inj_vs_detNL}, the orbital period and scaled planetary radius recovered with TLS after the filtering of the LCs of the quiet sample by N\&L are compared to the corresponding injected values, with different colors representing different planetary size. 

A possible explanation could reside in the TLS default template, which assumes circular orbits with a null impact parameter, as mentioned in Sect. \ref{sec:PLATO LCs}. Whenever the transit shape strongly differs from the default, the recovered scaled planetary radius may be underestimated with respect to the original injected value, while also affecting all the other related parameters, such as the semi-major axis and the transit duration. This occurs, for example, in the case of grazing transits (i.e., high $b$), eccentric orbits, or when the filtering algorithm distorts the original transit, as in the case of GPs or VARLET. 
This is a known issue of the VARLET algorithm, which however does not represent an obstacle as the search for transit signals is carried out by means of the BLS algorithm, which does not take the depth of the fitted box into account. 

Nevertheless, in  cases where the transit shape differs from the TLS default template, the correct parameters can be retrieved by performing an additional analysis in order to fit the correct transit model to the LC. For example, in the case of grazing transits or eccentric orbits, a Markov chain Monte Carlo (MCMC) analysis on the filtered flux is sufficient  to recover the orbital parameters correctly, as shown in Appendix A. When the orbital parameters are affected by a deformation of the transit shape, it is still possible to recover the original transit parameters by repeating the filter process after masking out the in-transit data points, which can be easily done after the transit has been detected. Therefore, we stress out the fact that the filtered LC used for the transit detection cannot be directly used for the planet characterization. An in-depth analysis is necessary for the precise measurement of the orbital and planetary parameters, either after reprocessing the LC while making use of the knowledge of the planet or by simultaneously fitting for stellar activity and planetary parameters.

\subsection{Entire light curves}\label{sec: entire_LC}
\begin{figure*}
  
\resizebox{\hsize}{!}
            {\includegraphics[width=10cm]{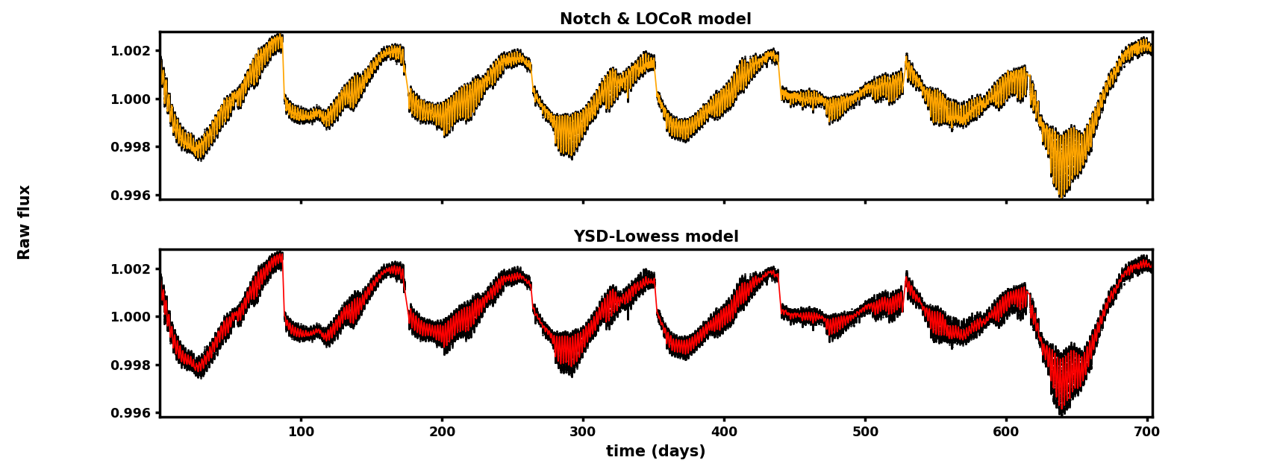}}
    \resizebox{\hsize}{!}
            {\includegraphics[width=10cm]{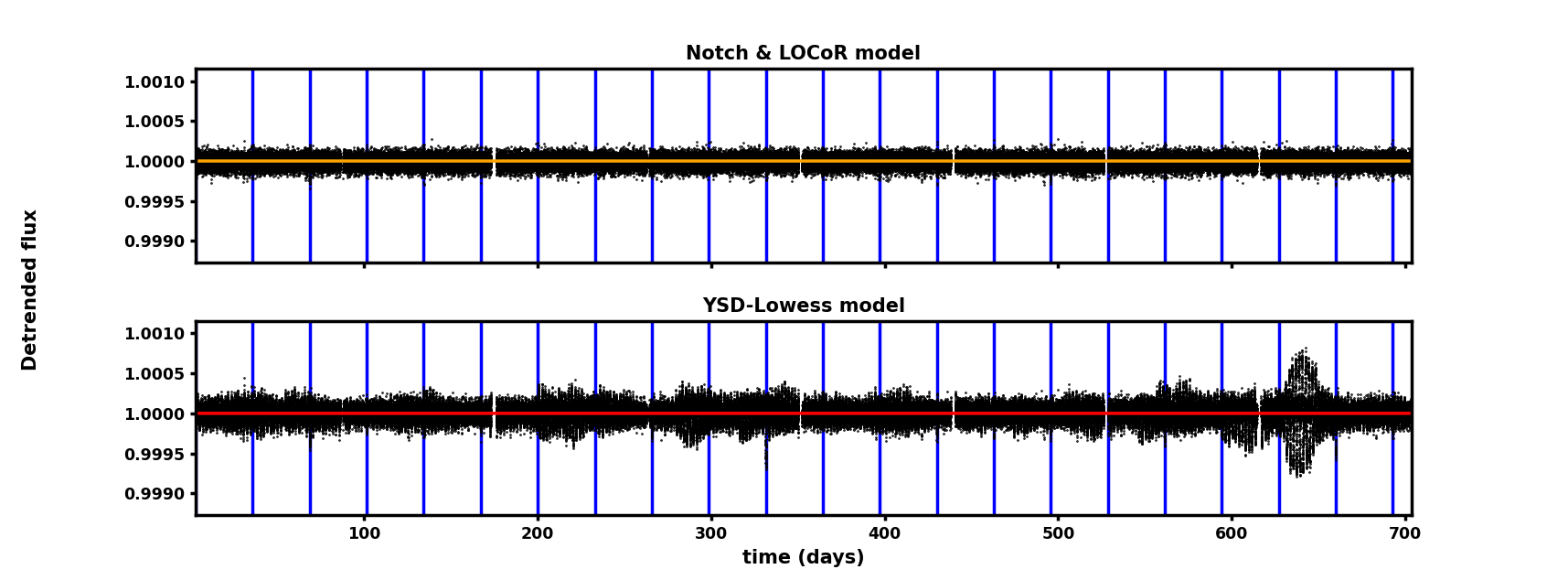}}
    \resizebox{\hsize}{!}
            {\includegraphics[width=10cm]{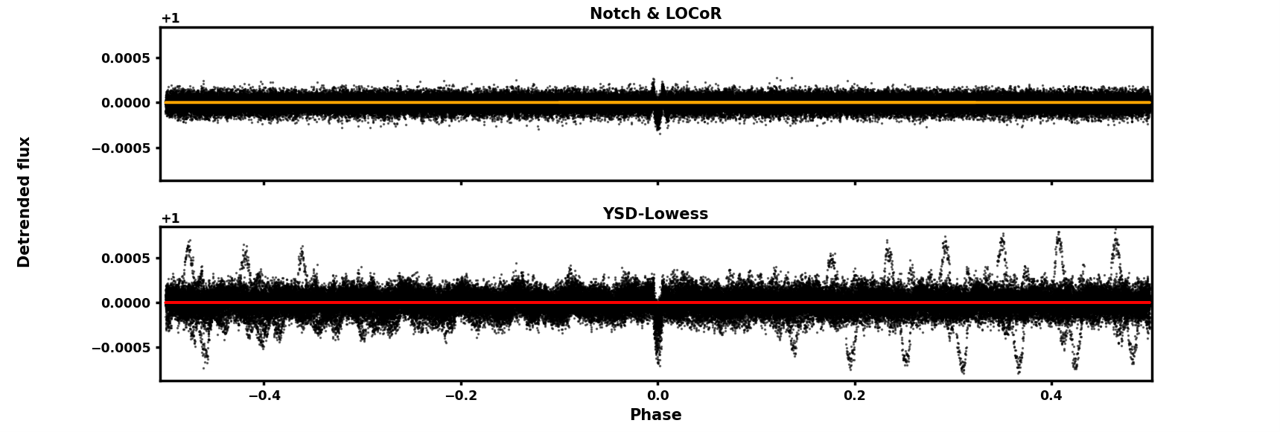}}

      \caption{Raw flux (upper panel), detrended flux (middle panel), and phase-folded flux (lower panel) of the LC \#58 of the sample of active stars with injected a transit signal with $P\mathrm{_{orb}}=32.83$ days and $R\mathrm{_p}/R_\star=0.015$. Data points are shown in black, whereas different colors represent the photometric variability model inferred from algorithm: N\&L (orange) and YSD (red). The transit is undetected in the former case, while it is detected with SDE=25 in the latter case. Blue vertical lines mark the mid-transit time.}
              
         \label{fig: LC158 allLC}
   \end{figure*}

One of the advantages of \textrm{PLATO} with respect to other space missions, such as \textit{TESS} and \textit{Kepler}, is the continuous observation of bright targets for at least two consecutive years, according to the current observation strategy. Such a long temporal baseline should allow to detect also the shallower transits of long-period small planets. 
In our previous analysis (i.e., Sect. \ref{sec: Hippke algorithms}, \ref{sec: individual_quarters} and 5.2), we analyzed only one quarter of data per planetary system, so it is possible that some small planets were not detected because there were not enough transits to allow for their detection, regardless of the filtering algorithm employed. To test this scenario, we selected the Earth-size planets (i.e., $R\mathrm{_p}/R_\star \leq 0.03$) left undiscovered after the N\&L, Huber spline, and YSD filtering (i.e., the three algorithms with the highest recovery efficiency; see Fig. \ref{fig: rec_fraction_active1}) and then injected their transits in the remaining seven quarters of the corresponding (active) LC. 
We then ran again the filtering algorithms and the TLS onto the entire LCs\footnote{The analysis on the entire LC was not performed for all transit signals because it was too computationally expensive.}. In this way, some of the previous missed transits are now detected whereas others are still missed, due to the extreme variability and high background noise of the active stars. Specifically, the percentage of recovered Earth-sized transits for N\&L increases up to 94.025\%, while for YSD, it reaches 96.55\% and for Huber spline, it is 98.075\%.\\
Surprisingly, the YSD and Huber spline algorithms seem to achieve better results than N\&L when considering a much larger dataset\footnote{A single quarter binned at 600 s has $\sim 12486$ data, whereas the entire LC has, on average, $\sim 99136$ points.}, recovering even some of the small planets still missed by N\&L.\\ An example is given in Fig. \ref{fig: LC158 allLC}, where the raw, detrended, and phase-folded fluxes of LC \#58 are shown in the upper, middle, and lower panels, respectively. The injected transit represents a small planet with a scaled planetary radius of $R\mathrm{_p}/R_\star=0.0146$ and an orbital period of $P\mathrm{_{orb}=32.83}$ days. In looking at the filtered fluxes (middle panel), it is quite hard to distinguish transit signals (marked by the vertical blue lines), especially in the case where the YSD model has been applied, due to a lot of residual noise left after the filtering. Nevertheless, when computing the phase-folded flux (bottom panel), in the N\&L case, the transit feature is barely distinguishable; in addition, it is so shallow that the TLS algorithm does not identify it as the strongest periodic signal, thus missing the planet. On the other hand, despite some residual variability, the transit feature is definitely recognizable in the phase-folded plot of the YSD case, with a much larger transit depth than the previous case and indeed it is correctly detected by TLS. 
This suggests that the algorithm efficiency could vary according to the number of quarters in a way that is not straightforward, thus it is recommended that two algorithms be used, instead of just one, when the dataset expands beyond one single quarter.

Finally, considering the entire LCs and the combined results of the three algorithms, we find that merging together the results of YSD and Huber spline increases the final detection efficiency close to 100\% for all planetary types, and specifically in Table \ref{tab: results} the final recovery percentages are reported.
\begin{table}[H]
\caption[]{Final recovery percentages of injected transit signals in simulated PLATO stellar LCs of a sample of active stars.}\label{tab: results} 
\centering
         \begin{tabular}{l c } 
            \hline
            \hline
            \noalign{\smallskip}
            Planetary type & recovered \%\\
            \noalign{\smallskip}
            \hline
            \noalign{\smallskip}
            Jupiter-size & 100\%\\
            Neptune-size & 99.9\%\\
            Earth-size & 98.875\%\\
            \noalign{\smallskip}
            \hline
            \hline \\
          \end{tabular}
     \end{table}
We also find that adding the Earth-size planets recovered by N\&L to those detected by YSD and Huber spline only slightly improves the final percentage of recovered transits, up to 99.0\%. Given that N\&L is much more computationally expensive than the other two algorithms, we do not recommend it to filter \textrm{PLATO} LCs encompassing several quarters. 
\subsection{Peculiar cases}
We investigated two peculiar cases that represent a small fraction of the existing planets but that are important from the planetary evolution point of view. The first one are eccentric hot Jupiters (described in Sect. \ref{sec: HJ}) and the second one is represented by those planets with an orbital period equal to the rotational period of their host star (reported in Sect. \ref{sec: Porb=Prot}).
We performed this analysis only on the LCs of the active sample and considering individual quarters, as in Sect. \ref{sec: individual_quarters}.

\subsubsection{Eccentric hot Jupiters}\label{sec: HJ}
\begin{figure}
\centering
\includegraphics[width=9cm]{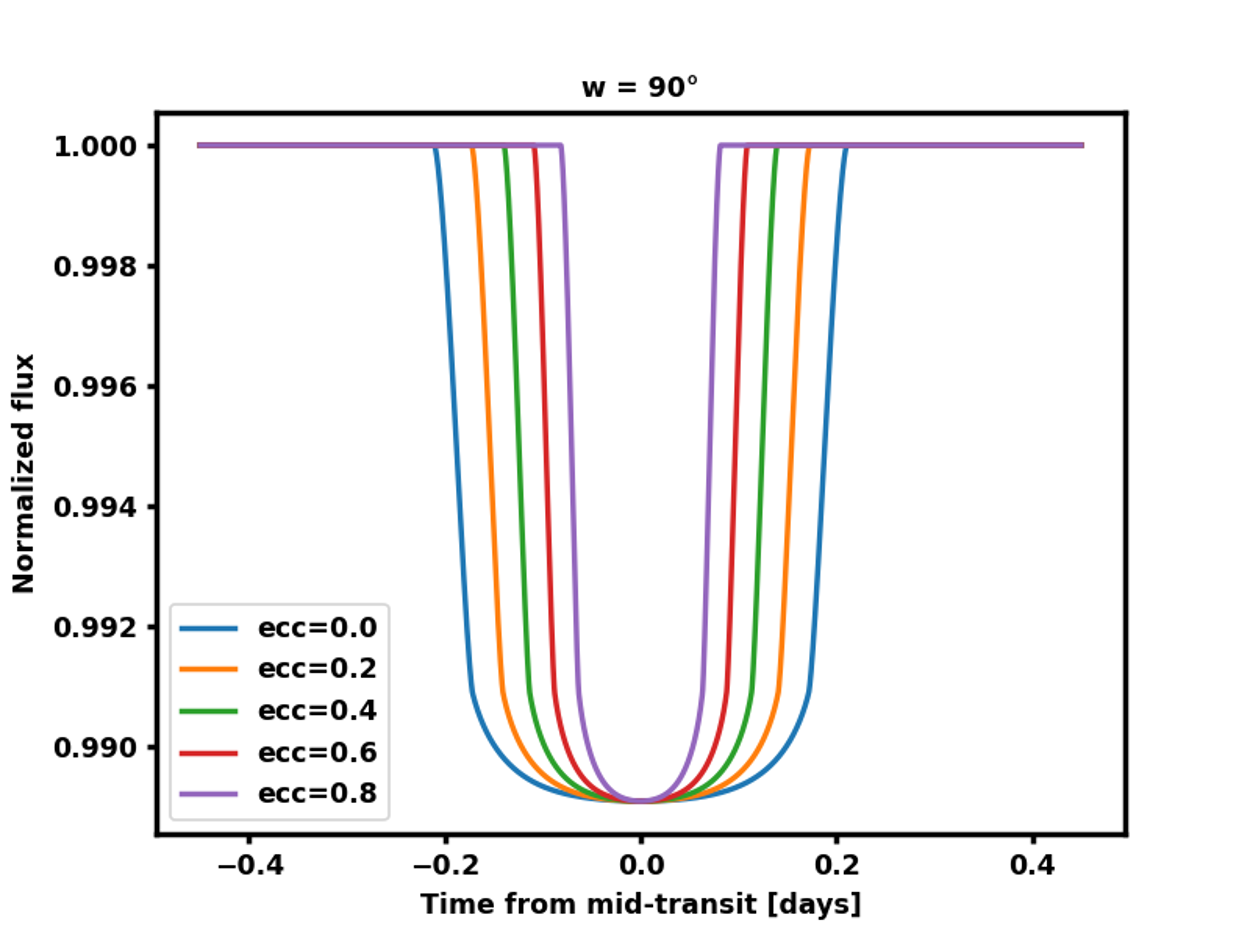}\hfil
\centering
\centering
\includegraphics[width=9cm]{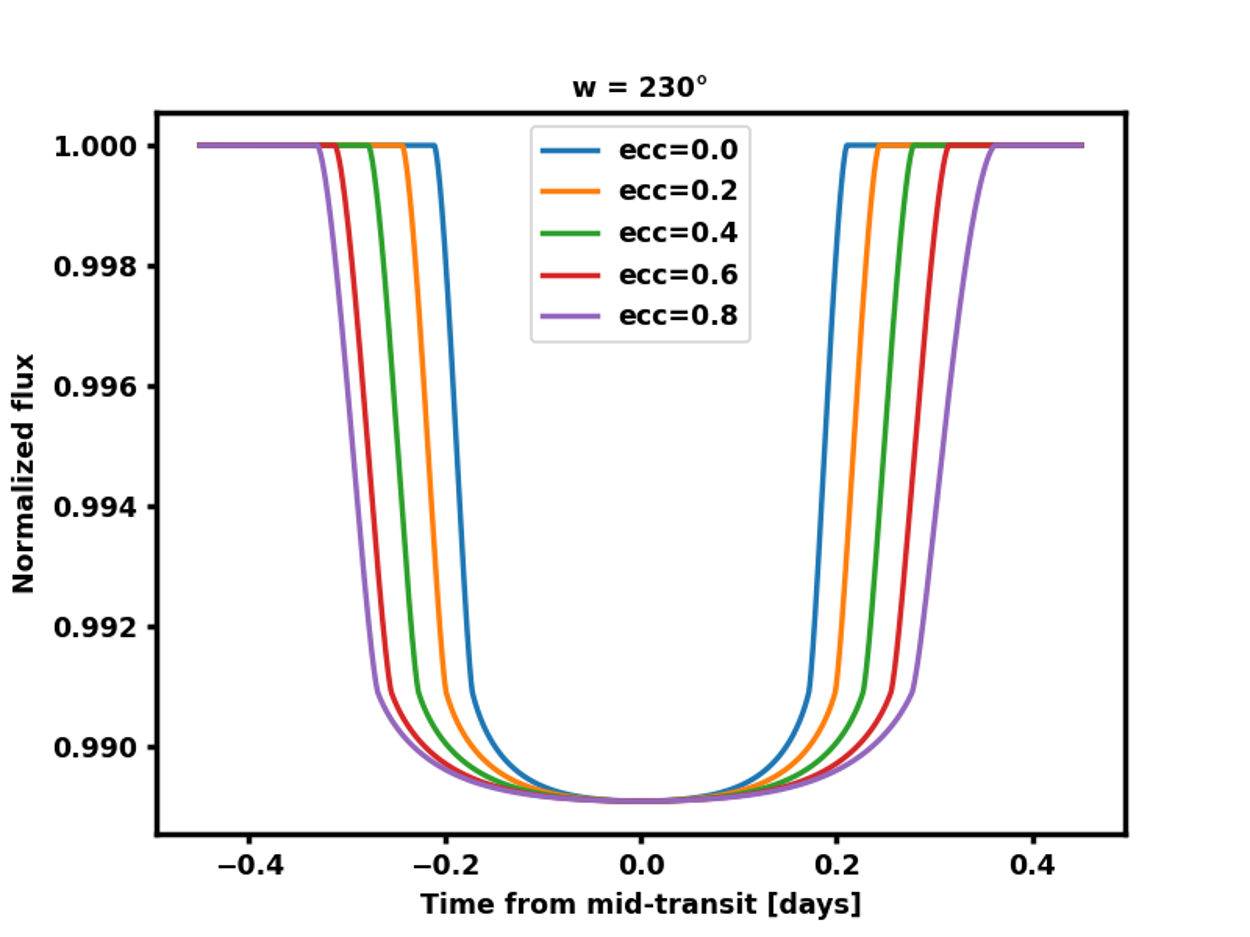}\hfil 
\centering
\caption{Illustration of transit shape variation at different orbital eccentricities and for two different argument of pericenter: $\omega=90^o$ (top) and $\omega=230^o$ (bottom).}\label{fig: varyingEcc}
\centering
\end{figure}

Hot Jupiters are large planets on extremely close-in orbits with periods shorter than 10 days. The formation pathways for these highly irradiated planets are still under investigation and among them, there are mainly two migration mechanisms: the so-called smooth migration and violent migration (\citealt{Bailey2018}). The latter scenario involves a series of high eccentricity trajectories as a consequence of gravitational interactions with the other bodies of the system, such as planet-planet scattering (\citealt{Beaug2012}) or the Kozai-Lidov mechanism (\citealt{Shevchenko2017}), until the planet is tidally captured onto a close-in circular orbit around the host star. Studying the detection efficiency of the filtering algorithms for these kind of transit signals is particularly interesting because determining if and how the frequency and the properties (such as the orbital eccentricity) of hot Jupiters vary as a function of age, allows us to understand which of the proposed evolutionary mechanisms actually occur. Therefore, we synthesized a set of 800 transit signals with the following properties: $R\mathrm{_p}/R_\star=0.1$, argument of pericenter ($\omega$) from a uniform distribution in the interval $[0-360]^o$, moving on orbits inclined to 90 degrees, with an eccentricity of 0.2, 0.4, 0.6, or 0.8 and an orbital period in the range of 1-10 days (again sampled from a uniform distribution). Indeed, eccentricity changes the transit shape, depending in turn on the argument of pericenter, as shown in Fig. \ref{fig: varyingEcc}, and we wanted to verify that also this kind of exoplanets could be detected. \\

We find that eccentricity does not really influence the detection efficiency; in fact, 100\% of injected signals were recovered after the filtering of all the selected algorithms. This result was partially expected since simulated transits belonging to this set represented all large planets, which are generally more easily detectable.
Besides the orbital periods and mid-transit times, which are also the reference parameters for the recovery of injected signals, the other recovered orbital and planetary parameters are often not in agreement with the corresponding injected values. This happens especially for planets in very eccentric orbits, when the transit fit performed with the TLS default template (with $ecc=0$) does not perfectly match the data-points. However, we verified that the correct orbital and planetary parameters can be retrieved by performing a MCMC fit. 
Moreover, in the VARLET case the algorithm reduces the transit depth of large planets, as already noticed in the previous tests. As a consequence, the SDE is decreased down to a maximum value of about 53, while the maximum SDE value of the recovered planets is around 74 for the other algorithms.  
\subsubsection{Stellar rotation-planetary orbit commensurability}\label{sec: Porb=Prot}
\begin{figure}[H]
   \centering
   \includegraphics[width=9cm]{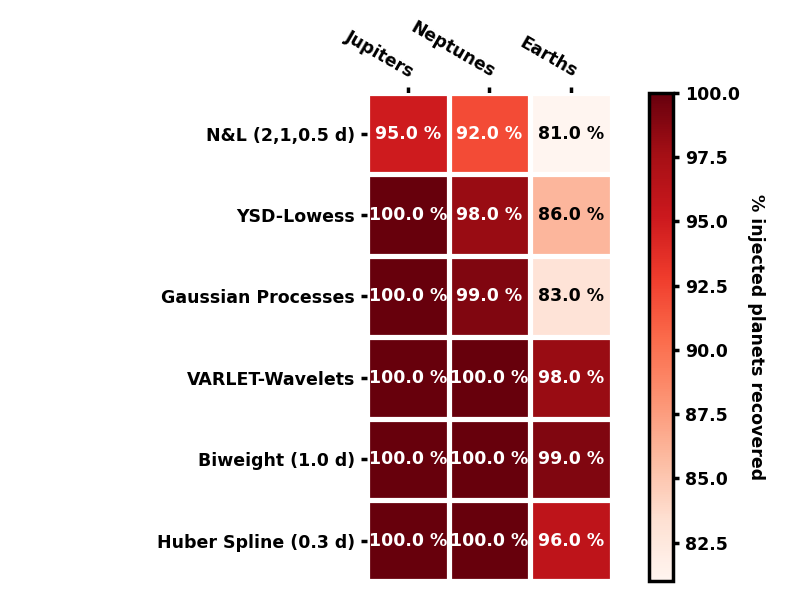}
      \caption{Recovery fraction of injected transit signals of planets with orbital period equal to the stellar rotation period in simulated \textrm{PLATO} stellar LCs of a sample of active stars.}
         \label{fig: rec_fraction_activeProt}
   \end{figure}

\begin{figure}[H]
\centering
\includegraphics[width=9cm]{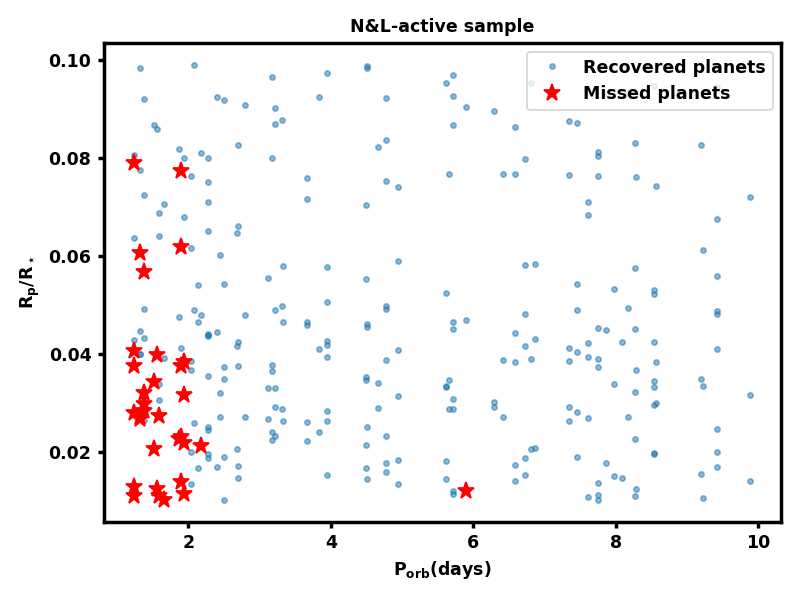}\hfil
\centering
\caption{Injection-recovery tests on the sample of active stars for planetary transit signals with $P\mathrm{_{orb,planet}}=P\mathrm{_{rot,star}}$. Injected scaled planetary radius as a function of the injected orbital period: blue dots are the recovered planets, whereas red stars are the missed planets after filtering the LCs with Notch \& LOCoR.
}\label{fig: Rp_Rs vs Porb active5} 
\centering
\end{figure}

The second peculiar case is characterized by those planets which orbit with a period exactly equal to their host star rotation period, namely $P_\mathrm{orb,planet}=P_\mathrm{rot,star}$. Several stars are observed hosting close exoplanets (mostly giant planets) rotating in an orbit with a period equal to an integer multiple of the stellar rotation period (\citealt{Walker2008}; \citealt{Pagano2009}; \citealt{Szabo2012}; \citealt{Beki2014}). This is reflected in the photometric observations of the host star which  exhibits variations synchronous to the orbital period of the corresponding planet. Studying exoplanetary systems with a stellar rotation-planetary orbit commensurability is important in order to better understand how interactions between the host star and the planet work. Nevertheless, it is challenging to find transiting exoplanets whose orbital period closely matches the rotation period of the star or is commensurable to it, since the transit signal feature may be indistinguishable from the rotational variability of the host star.\\
This case was tested in this work only on the first quarter of each LC of the active sample. We generated a library of 300 different template transits assuming a circular orbit with an impact parameter $b$ uniformly sampled in [0,1], and with a planet-to-star radius randomly drawn in the range  $R\mathrm{_p}/R_\star \in [0.01-0.1]$, dividing the set into three planetary types depending on $R\mathrm{_p}/R_\star$, as previously done (see Sect. \ref{sec: tests}): Jupiter-, Neptune-, and Earth-size planets. Specifically, we produced 100 transit templates with planet-to-star radius in [0.01-0.03] (i.e., Earth-size), 100 with $R\mathrm{_p}/R_\star \in [0.03-0.05]$ (i.e., Neptune-size), and 100 with $R\mathrm{_p}/R_\star \in [0.05-0.1]$ (i.e., Jupiter-size). For all these planets, we chose the orbital period to be exactly equal to the rotation period of their host star.\\
The percentage of recovered transits of different planetary type is shown in Fig. \ref{fig: rec_fraction_activeProt}. Since the simulated LCs of the active sample have $P\mathrm{_{rot} \leq 10}$ days, there are many injected transits representing ultra-short-period planets, meaning a category of exoplanets defined by having orbital period shorter than a single day. This kind of planet is not so rare, they are about as common as hot Jupiters. Indeed it is estimated that about one out of 200 Sun-like stars has an ultra-short-period orbiting planet (\citealt{Winn2018}).
 \vspace{-10pt}
From Fig. \ref{fig: rec_fraction_activeProt}, it can be noticed that, generally, all the filtering algorithms (except N\&L) show a good performance for Jupiter- and Neptune-size planets, with recovery efficiency higher than 98\%,  while in the case of Earth-size planet only biweight and VARLET reach this threshold. 
Indeed, N\&L under-performs with respect to the other algorithms, recovering only 89.33\% of the total planets. The missed planets are those with an orbital period shorter than 2 days, namely, those for which also the \texttt{LOCoR} filtering method is used (as shown in Fig. \ref{fig: Rp_Rs vs Porb active5}), where the scaled planetary radius as a function of the injected orbital period is shown. 
This result was partially expected since also \citet{Rizzuto2017} (authors of this code) noticed that the \texttt{LOCoR} algorithm, which is applied to the most rapidly rotating stars (i.e., $P\mathrm{_{rot}\leq 2}$ days), tends to fail in modeling the photometric variability without affecting the in-transit points when the planetary orbital period closely matches the stellar rotation period.
The general-purpose algorithms perform very well for this kind of planets, specifically, the biweight with a window-size of 1 day shows the highest detection efficiency, recovering up to 99\% of injected transits of small planets. The Huber spline shows a very high recovery efficiency as well, although it is lower than that of VARLET.\\ 
Among the custom-built algorithms, VARLET shows the highest recovery percentage, detecting 99.33\% of the total injected transits, and, in particular, 98\% of the small planets, a surprising result when compared to those detailed in Sect. \ref{sec: individual_quarters}. 
Although this algorithm achieves a detection efficiency of 100\% for larger planets, it strongly reduces their transit depth, as already observed, recovering them with a low SDE of about $\sim$54. For larger planets similar performances are achieved by YSD, where planets are recovered with a higher SDE (up to 90) than VARLET, although some are missed.\\

\begin{figure}[H]
   \centering
   \includegraphics[width=9cm]{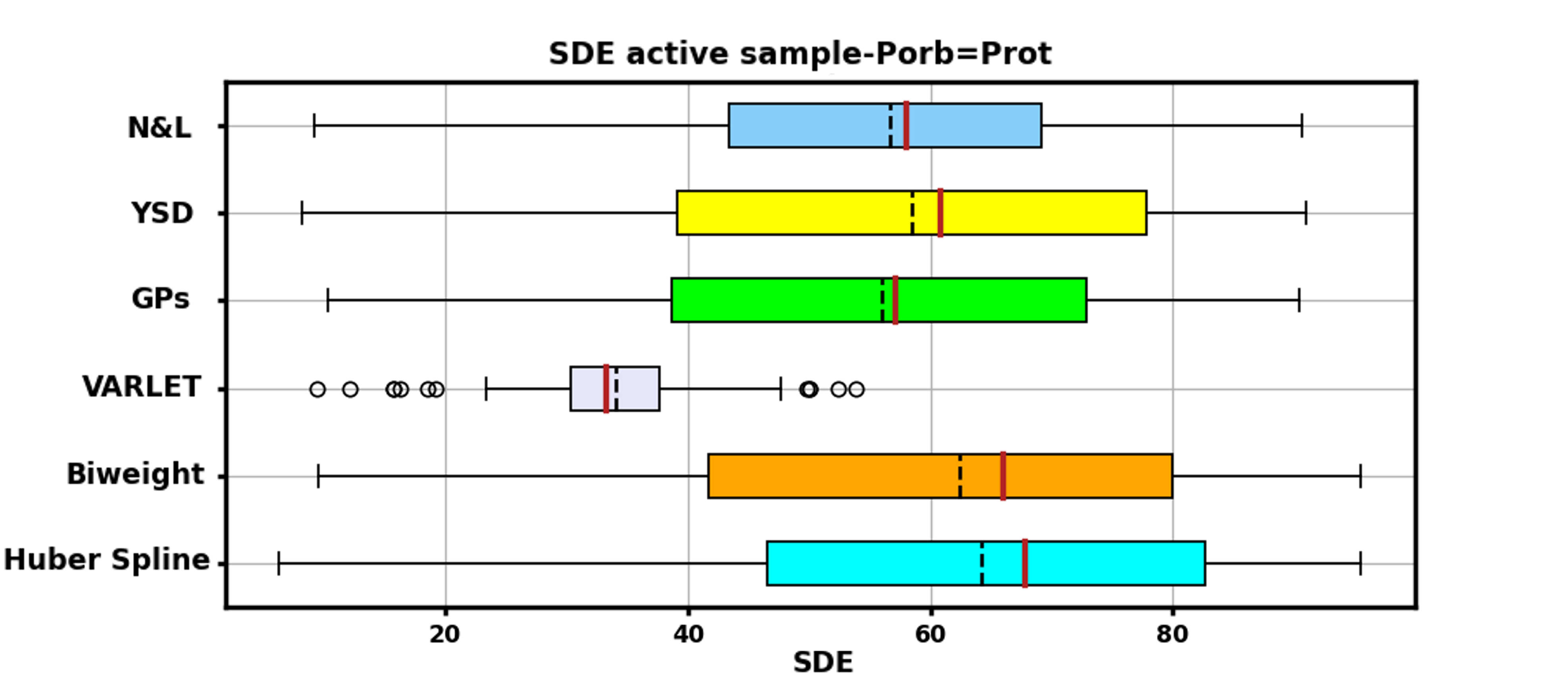}  
      \caption{Boxplot of the SDE for the algorithms listed on the y-axis, tested on the active sample with injected transit signals with $P_\mathrm{orb,planet}=P_\mathrm{rot,star}$. Boxes cover the lower to upper quartiles; whiskers show the 10 and 90 percentiles. Dashed black lines indicate the mean SDE, whereas red lines show the median. The black circles in the VARLET case represent data that extend beyond the whiskers.}\label{fig: SDE_active2}
   \end{figure}

In Fig. \ref{fig: SDE_active2} the SDE boxplot summarizing the properties of the SDE distributions is shown. With regards to the SDE values of YSD and biweight (yellow and orange bars, respectively) they show very similar results, even if the biweight median SDE is slightly larger than the median of YSD. Nevertheless, the highest median SDE is obtained by the Huber spline algorithm, which, however exhibits slightly worse performances than the other two algorithms.
   \vspace{-10pt}
\section{Discussions and conclusions}\label{sec: conclusions}
The idea behind this work was inspired by the \textrm{PLATO} Working Package (WP) 111 000 (\textit{Coordination of Tools for Lightcurve Filtering}), a sub-group of the WP 11 that deals with \textrm{PLATO's} \textit{Exoplanet Science}. In preparation for this space mission, it was necessary to determine which is the best-performing algorithm for the filtering of stellar activity in \textrm{PLATO} LCs, for the subsequent detection of planetary transit signals. The results presented in this work suggest that the best strategy to adopt when analysing \textrm{PLATO} LCs should be as follows:
\begin{itemize}
    \item Using the N\&L algorithm to filter activity on LCs obtained every three months (i.e., each time that a new quarter of data is added to the previous time series), since it turns out to be the best-performing algorithm on partial time series for both the quiet and active sample. Alternatively, similar results can be achieved by combining the Huber spline and the biweight algorithms.
    \item After two years, at the end of the observations, that is, once that the LCs of the targets are completed, using both Huber spline and YSD to filter stellar activity on the entire LC (i.e., all the 8 quarters). Indeed, from this analysis, it appears that the combination of these two algorithms succeeds in detecting small planets that are missed by N\&L on long temporal baselines, thus achieving the highest possible efficiency.
    \item For stars with a rotation period shorter than 2 days\footnote{The stellar rotation period will be determined by the WP 12, which is in charge of \textrm{PLATO} \textit{Stellar Science}, prior to the transit search.}, it is recommended to use the biweight algorithm on the entire LCs, since it shows the highest recovery efficiency and the highest SDE values in detecting small exoplanets with an orbital period that closely matches the stellar rotation period. This kind of transit signals are often missed by N\&L. We note that in this case, VARLET is the second best-performing algorithm despite the lower average SDE, in opposition to the general trend observed in the other cases. 
\end{itemize}
Among the tested algorithms, the GPs showed poor performances compared to the others. However, we would like to emphasize that we investigated Gaussian Processes using only a specific kernel (the Celerite rotation term implemented in \texttt{celerite2}) and that the results could significantly change with a different kernel, such as a Matern 3/2 term or a squared exponential. It is well known that Gaussian processes tend to absorb the transit signal, by modeling it as part of the photometric variability, and that this problem can be alleviated by simultaneously fitting for the planetary transit.
We also note that the biweight algorithm can achieve better performances by fine-tuning the window width to three times the transit duration, as found by H19, although in our case, there is no prior knowledge of the orbital parameters of the planet. In our analysis, we investigated the biweight method using a fixed window of 1.0 day as the best approximation given the expected maximum transit duration of our injected transit models; however, possible performance gains could be achieved with a different value or by applying an adaptive window width.\\ \\
In a recent paper, \citet{Nardiello2021} investigated the stellar age versus planetary radius distribution by using the results of different works on both open clusters and young stellar associations, with well-constrained stellar ages, containing stars hosting both confirmed and candidate exoplanets. This distribution is shown in Fig. 7 of \citet{Nardiello2021}, where it is clearly visible that small  exoplanets, namely, objects with $R\mathrm{_p} \leq 4 R\mathrm{_{Earth}}$, are all concentrated at ages older than 100 -- 200 Myr, whereas larger planets exhibit a broader distribution. This concentration of small objects around older stars might be due to either planetary formation and evolution processes or to an observational bias, since  it has been particularly difficult to detect transits of Earth and Super-Earth-size planets in the LC of young active stars to date. 
Until a systematic analysis over all ranges of planetary radius is performed, it is difficult to draw any theoretical conclusion with certainty. To date, such  an analysis cannot be executed with current astrophysical instrumentation, but it may be possible thanks to \textrm{PLATO} in the coming years. Indeed, according to the results of this work, it seems that when turning from \textit{TESS} to \textrm{PLATO} LCs of young active stars, the efficiency of the filtering algorithms for the planetary transit detection increases from about 40\% up to nearly 99\%. 
The injection-retrieval experiments performed on \textrm{PLATO} simulated LCs have shown that \textrm{PLATO} will be able to detect a large variety of exoplanets with many different characteristics, in terms of size, orbital periods, and eccentricity. In particular, they have demonstrated that detecting Earth-size exoplanets with \textrm{PLATO} will be possible with both general-purpose and custom-built filtering algorithms, even in stellar LCs of young active stars. Therefore, in the forthcoming years only \textrm{PLATO} can tell us if the lack of small exoplanets around young stars is really related to an observational bias.\\
Even if the injection-recovery tests performed in this work included several different kinds of planets, we have not explored yet the entire parameter space of the observed exoplanets. Considering the distribution of planets detected by the transit technique, we are far from completion. Other peculiar cases that could be tested and simulated in future works include, for example, transits showing timing variations (TTV, \citealt{Holman2005}, \citealt{Nascimbeni2011}, \citealt{Malavolta2017}), monotransits (\citealt{Cooke2020}), multi-planetary systems, or small planets with orbital periods longer than 40 days, that is, temperate terrestrial planets such as those around the dwarf star TRAPPIST-1 (\citealt{Gillon2017}). Indeed, \textrm{PLATO}'s top-level science requirements are focused on habitable planets, therefore it would be interesting to test the algorithms also on more temperate Earth-like planets. Efforts to compute the expected yield of Earth analogs by \textrm{PLATO} are already ongoing,  for example, by \cite{Heller2022}.\\
Furthermore, in future works, other filtering algorithms can be tested in order to analyze their efficiency on \textrm{PLATO} LCs and to compare the results with those presented in this paper. For example, the Kepler Science Data Processing pipeline\footnote{https://github.com/nasa/kepler-pipeline} (\citealt{Jenkins2010}) that is now publicly available, has not been tested on active stars yet, but it has demonstrated very good results on quiet stars, so it would be interesting to check its performance on the active sample of \textrm{PLATO} LCs. \\
Finally, we note that for our experiment, we employed the TLS algorithm only, assuming that a transit-like shape could be more efficient than a box-like shape in detecting the planet. However, the deformation of the transit feature introduced by a filtering algorithm may invalidate this assumption (see Fig. \ref{fig: LC22 zoom}). Future works related to the planetary yield of \textrm{PLATO} should consider multiple filtering algorithms as well as multiple detection methods. 
It is also worth noting that the full \textrm{PLATO} pipeline is now under development, as well as extended simulations including the effects of contaminants in the LC, while a complete and robust assessment of the in-flight performance will be available only after the launch. Although we could have removed the long-term instrumental systematics; for instance, with the use of a Gaussian Process, we preferred to use the LC as they are produced by the \textrm{PLATO} Solar-like Light-curve Simulator and test the worst-case scenario in which an intermediate data product has to be analyzed. As our analysis is purely differential, that is, we are not trying to determine \textrm{PLATO}'s yield for a specific class of planets, our results should hold as long as the comparison is performed in a self-consistent way and the main properties of the simulations resemble those of the final LCs. When simulated final \textrm{PLATO} data products  become available, we plan to perform new injection-retrieval tests, including both peculiar cases not considered in this work, such as mono-transits or long-period planets, as well as custom-built filtering algorithms that will be presented to the community.

\begin{acknowledgements} 
This work presents results from the European Space Agency (ESA) space mission
PLATO. The PLATO payload, the PLATO Ground Segment and PLATO data processing
are joint developments of ESA and the PLATO Mission Consortium (PMC). Funding for
the PMC is provided at national levels, in particular by countries participating in the
PLATO Multilateral Agreement (Austria, Belgium, Czech Republic, Denmark, France,
Germany, Italy, Netherlands, Portugal, Spain, Sweden, Switzerland, Norway, and United
Kingdom) and institutions from Brazil. Members of the PLATO Consortium can be found
at \url{https://platomission.com/}. The ESA PLATO mission website is
\url{https://www.cosmos.esa.int/plato}. We thank the teams working for PLATO for all their work. L.M., I.P., G.P., and S.D. acknowledge support from PLATO ASI-INAF agreements n.2015-019-R0-2015 and n. 2015-019-R.1-2018.
\end{acknowledgements}

\bibliographystyle{aa}
\bibliography{references} 


\begin{appendix} 
\section{Orbital and planetary parameters from TLS and MCMC fit of LC \#17}
\begin{figure*}
  \centering
   \resizebox{15cm}{!}
            {\includegraphics[width=1cm]{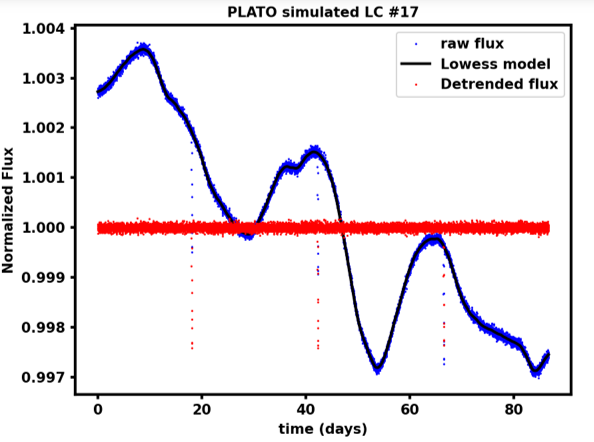}}
    \caption{Filtering by YSD of the first quarter of simulated LC \#17 of the quiet sample with injected a large planet. In blue, we show the raw flux, in black the filtering model, and in red the corrected flux are shown, as in the legend.}\label{fig: LC17 detrending}          
   \end{figure*}

\begin{figure*}
\resizebox{\hsize}{!}
{\includegraphics[width=10cm]{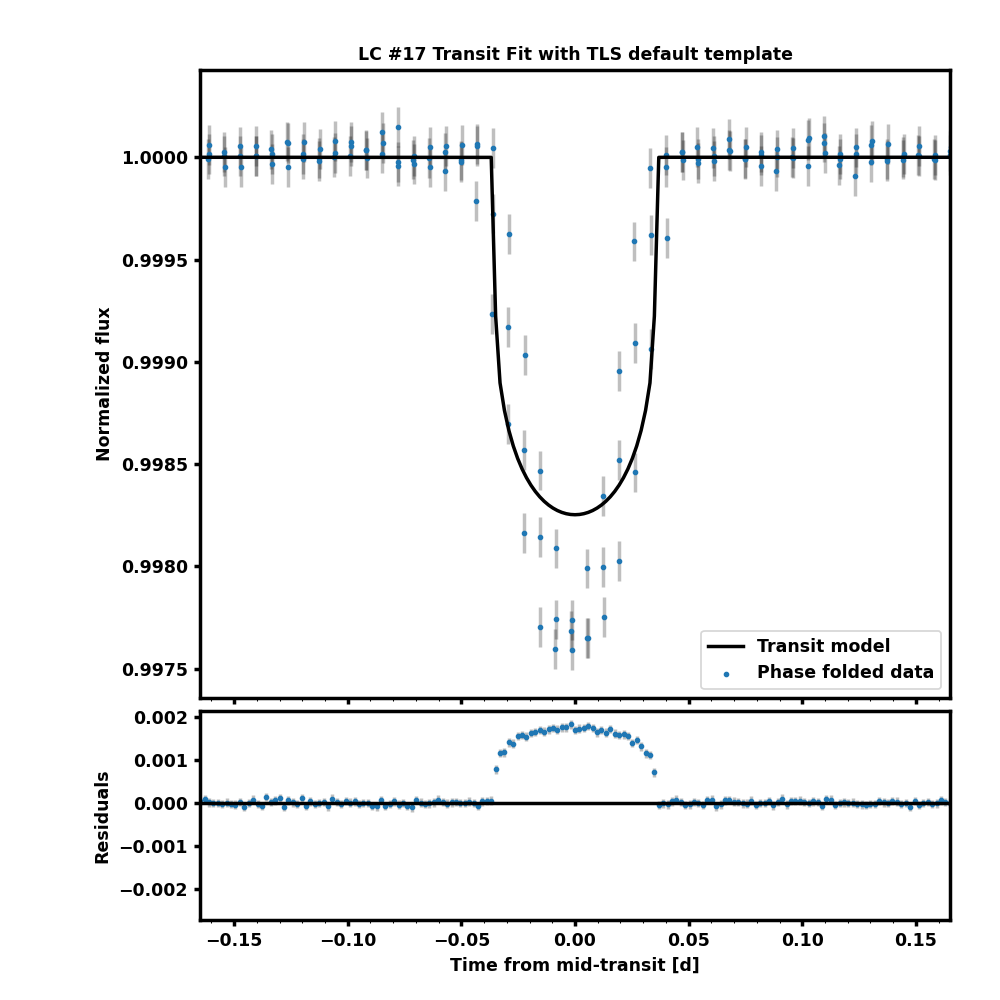}
\includegraphics[width=10cm]{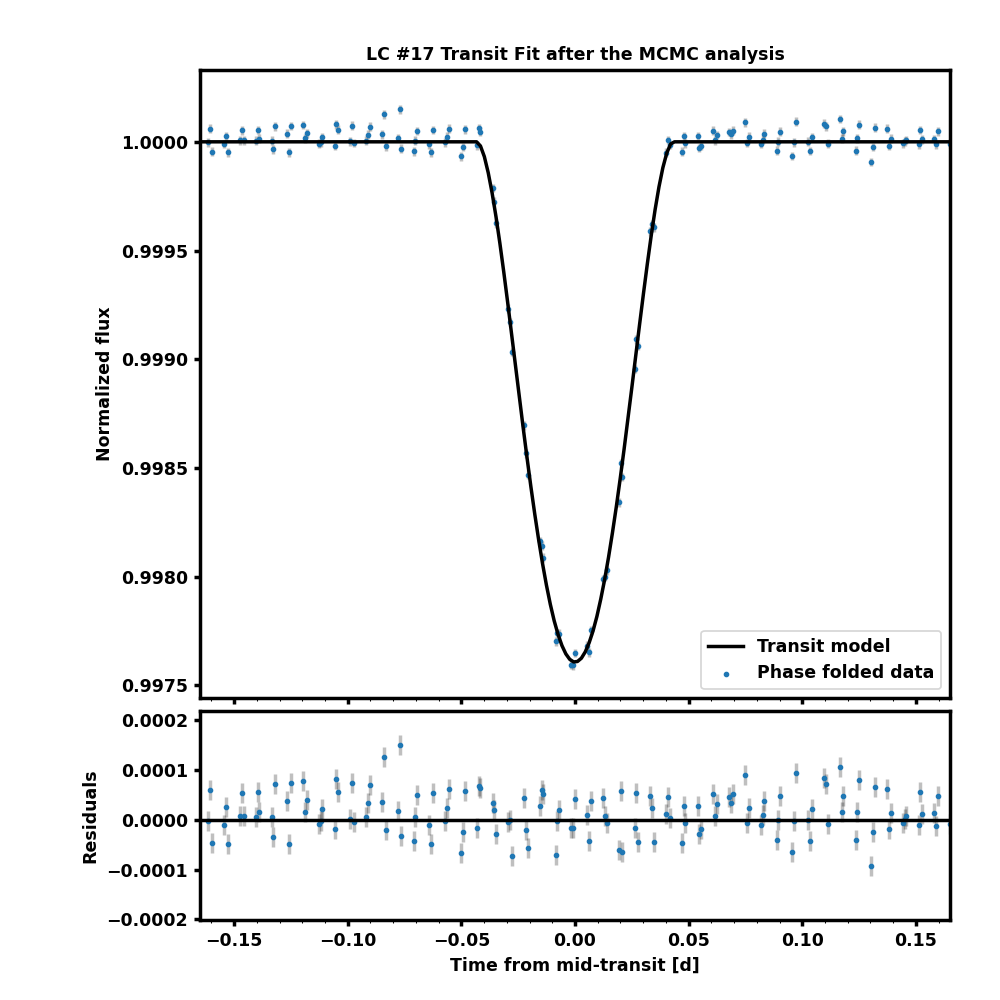}}
\caption{Transit fit of the first quarter of LC \#17 with a large  injected planet. Top panel:\ Best fit of the transit light curve is represented by the solid black line, whereas the blue points are the \textrm{PLATO} normalized data, as a function of the time from mid-transit. Bottom panel:\ Residuals (O-C: observed data-calculated data) are shown as a function of the time from mid-transit. Left: Transit model computed with TLS default template. Right: Transit model computed after an MCMC analysis.}\label{fig: LC17 transit fit}
\centering
\end{figure*}

An example of a grazing transit when the TLS default template does not correctly fit the data-points (and, consequently, it fails to retrieve most of the orbital and planetary parameters) is given in Fig. \ref{fig: LC17 transit fit}. This transit represents a large planet with a $R\mathrm{_p}/R_\star \sim 0.09$, whereas the TLS recovered a planetary signal corresponding to a much smaller body of only $R\mathrm{_p}/R_\star \sim 0.037$, as shown in Fig. \ref{fig: LC17 transit fit} (left), where the phase-folded data and the residuals of the transit fit performed with TLS are shown in the upper and lower panel, respectively. In Fig. \ref{fig: LC17 detrending} the filtering performed by YSD on the raw flux is shown. Even when performing a fit with the TLS grazing template, the inferred scaled planetary radius is 0.043, slightly improving with respect to the TLS default template fit but still underestimating the injected value. Therefore, a Markov chain Monte Carlo (MCMC) analysis has been run in order to fit the correct transit model to the light curve and thus achieve a more accurate planet characterization. For this purpose, the \texttt{PyORBIT}\footnote{https://github.com/LucaMalavolta/PyORBIT} code (\citealt{Malavolta2016}) was used to perform a MCMC analysis with 16 walkers and 500000 steps (the burn-in is after 20000 steps), initialized with a normal distribution around stellar radius and density. Moreover, uniform priors for the other parameters have been applied, such as the period and the time of transit, centered on the values obtained with TLS, within their boundaries. We further assumed, as priors on the limb darkening coefficients $u_1$ and $u_2$, a Gaussian distribution centered on the values given by tables describing the simulated LCs, provided by LSWG. The relevant physical parameters obtained from the posterior samples of the best-fit results are reported in Table \ref{tab: LC17 derived parameters}.\\
Uncertainties on the orbital and planetary parameters were inferred from their a posteriori distributions obtained from the MCMC analysis. 
The transit fit and its residuals are shown in Fig. \ref{fig: LC17 transit fit} (right). It is possible to notice that the transit fit significantly improves by performing the MCMC analysis, recovering parameter values consistent with the injected ones. This suggests that the scaled planetary radius obtained with TLS must be considered carefully, since the real transit shape could significantly differ from the TLS template, thus resulting in a systematic error which gives an under-estimate of $R\mathrm{_p}/R_\star$ -- consequently affecting also all the other parameters related to $R\mathrm{_p}/R_\star$. On the other hand, the estimates of the orbital period and the time of transit recovered from the TLS transit search result to be very precise and generally reliable.\\ Finally, we emphasize that the TLS algorithm has not been designed for transit characterization, but only for transit detection. Indeed, an MCMC analysis generally needs to be performed in order to characterize real planetary transits. 

\begin{table}[H]
\caption[]{LC \#17: Orbital and planetary parameters injected and derived from the first quarter of \textrm{PLATO} simulated LC \#17, performing a transit fit with TLS and PyORBIT.}\label{tab: LC17 derived parameters} 
         \begin{tabularx}{0.5\textwidth}{l c c c}
            \hline
            \hline
            \noalign{\smallskip}
            Parameter & Injected value & Recovered with & Recovered with\\
             & & TLS & MCMC \\
            \noalign{\smallskip}
            \hline
            \noalign{\smallskip}
            $P\mathrm{_{orb}}$(days) & 24.2286 & 24.222 $\pm$ 0.030 & 24.2284 $\pm$ 0.0002 \\
            $T_0$ (days) & 18.0648 & 18.0720 & 18.0649 $\pm$ 0.0002 \\
            $b$  & 0.975 & 0 (default) & 0.97 $\mathrm{^{+0.07}_{-0.02}}$\\
            $R\mathrm{_p}/R_\star$ & 0.090 & 0.037 & 0.089 $\mathrm{^{+0.056}_{-0.012}}$ \\  
            $\rho_\star$(g cm$^{-3}$) & 2.68 & - & 2.68 $\mathrm{^{+0.31}_{-0.46}}$\\
            $u_1$ & 1.129 & 0.4804 (default) & 1.129 $\pm$ 0.005\\
            $u_2$ & -0.169 & 0.1867 (default) &  -0.170 $\pm$ 0.005\\
            jitter & - & - & (4.35 $\pm$ 0.06)$\cdot 10^{-5}$\\          
            $a/R_\star$ & 43.678 & 100.621 & 43.623 $\mathrm{^{+1.65}_{-2.68}}$ \\
            $i$ (deg) & 88.72 & 90 (default) & 88.72 $\mathrm{^{+0.07}_{-0.18}}$ \\
            $R\mathrm{_p} (R\mathrm{_j})$ $^{(a)}$ & - & 0.4720 &  1.14 $\mathrm{^{+0.83}_{-0.19}}$ \\ 
            $T_{14}$ (days)$^{(b)}$ & - & 0.0766 & 0.0864 $\mathrm{^{+0.00010}_{-0.0009}}$ \\
            \textit{a} (AU) $^{(a)}$ & - & 0.6124 & 0.2665 $\pm$ 0.0004 \\ 
            $e^{(c)}$ & 0 & 0 & 0 \\
            \noalign{\smallskip}
            \hline
            \hline \\
          \end{tabularx}
         
         \textbf{Notes:}\\
         $^{(a)}$ $R\mathrm{_p} (R\mathrm{_j})$ and \textit{a}(AU) are computed assuming a stellar radius of $R_\star=1.311 R_\odot$, derived from PARSEC (\citealt{Bressan2012});\\
         $^{(b)}$ $T_{14}$: total transit duration, time between first and last contact;\\
         $^{(c)}$ The orbit is assumed to be circular, hence the eccentricity is fixed to 0.
         
     \end{table}
     
\section{Stellar parameters of the simulated LCs}\label{app: stars}
\begin{figure}[H]
\centering
\includegraphics[width=.6\textwidth, center, right]{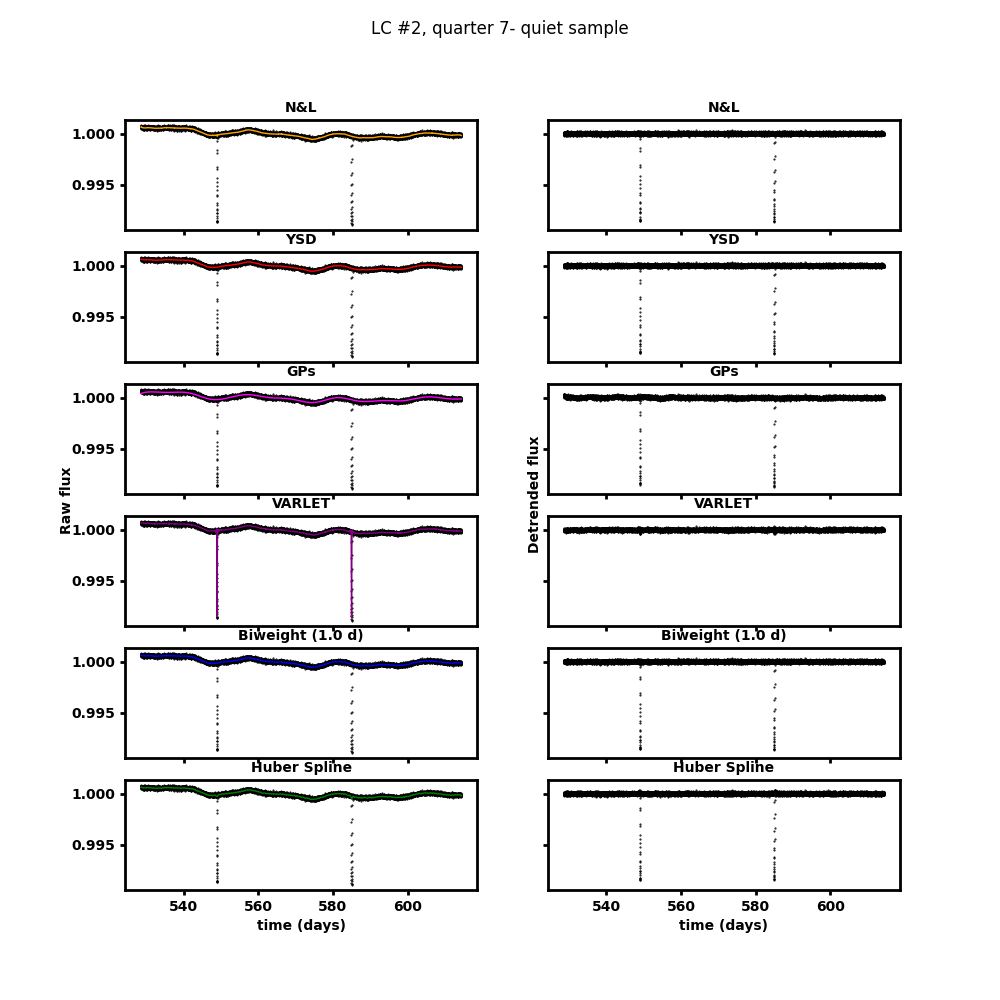}\hfil
\centering
\vspace{-15pt}
\caption{Filtering of quarter 7 of LC \#2 of the quiet sample with an injected large planet. The filtering models are overplotted in different colors and the resulting detrended flux is shown on the right panels.}\label{fig: LC2q7_quiet} 
\centering
\end{figure}

\begin{figure}[H]
\raggedright
\centering
\includegraphics[width=.6\textwidth, center, right]{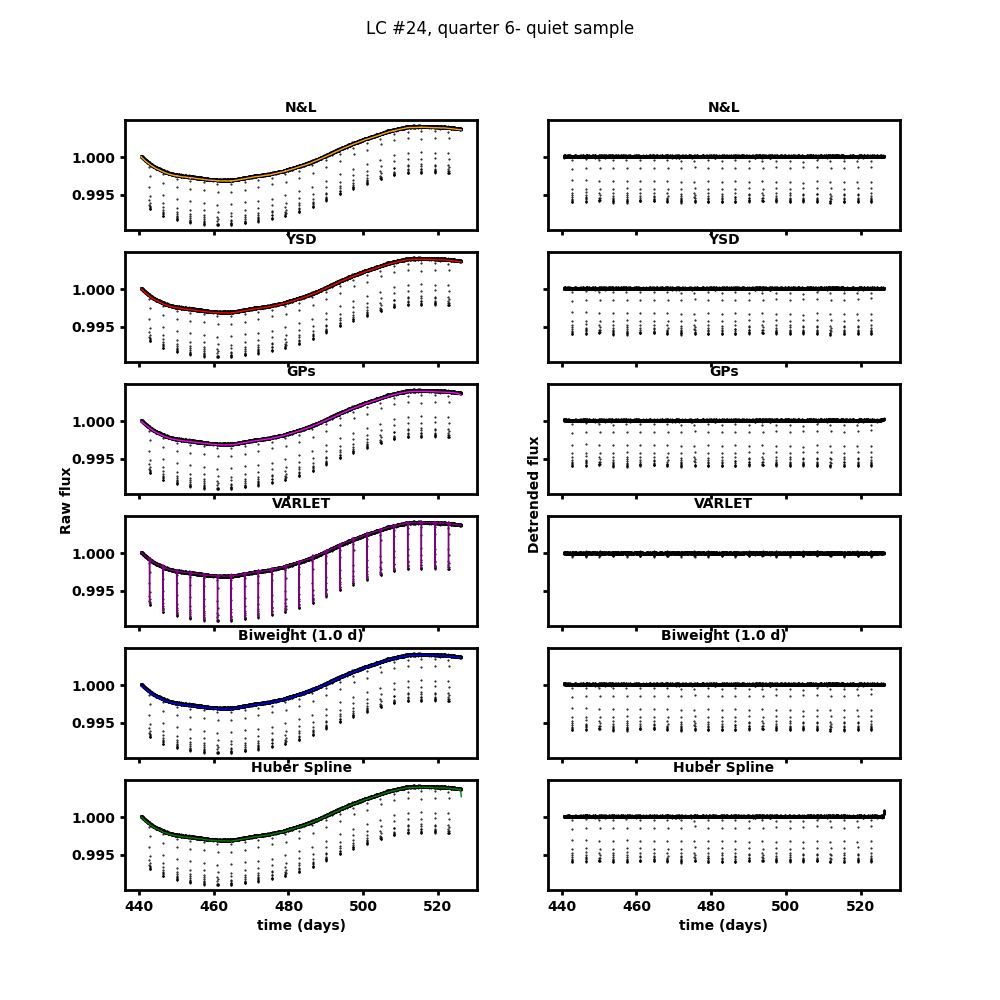}\hfil
\centering
\vspace{-15pt}
\caption{Filtering of quarter 6 of LC \#24 of the quiet sample with an injected large planet. The filtering models are overplotted in different colors and the resulting detrended flux is shown on the right panels.
}\label{fig: LC24q6_quiet} 
\centering
\end{figure}

\begin{figure}[H]
\centering
\includegraphics[width=.6\textwidth]{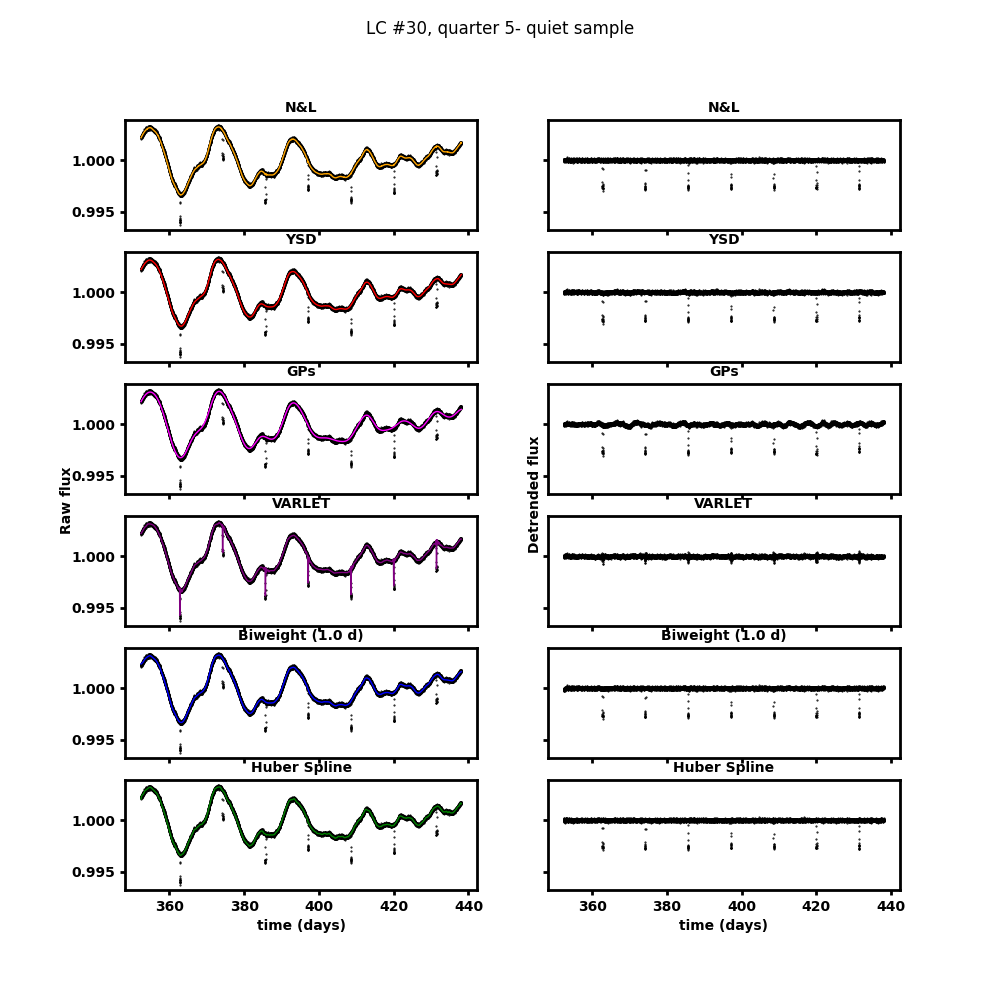}\hfil
\centering
\vspace{-25pt}
\caption{Filtering of quarter 5 of LC \#30 of the quiet sample with an injected large planet. The filtering models are overplotted in different colors and the resulting detrended flux is shown on the right panels.
}\label{fig: LC30q5_quiet} 
\centering
\end{figure}

\begin{figure}[H]
\centering
\includegraphics[width=.6\textwidth]{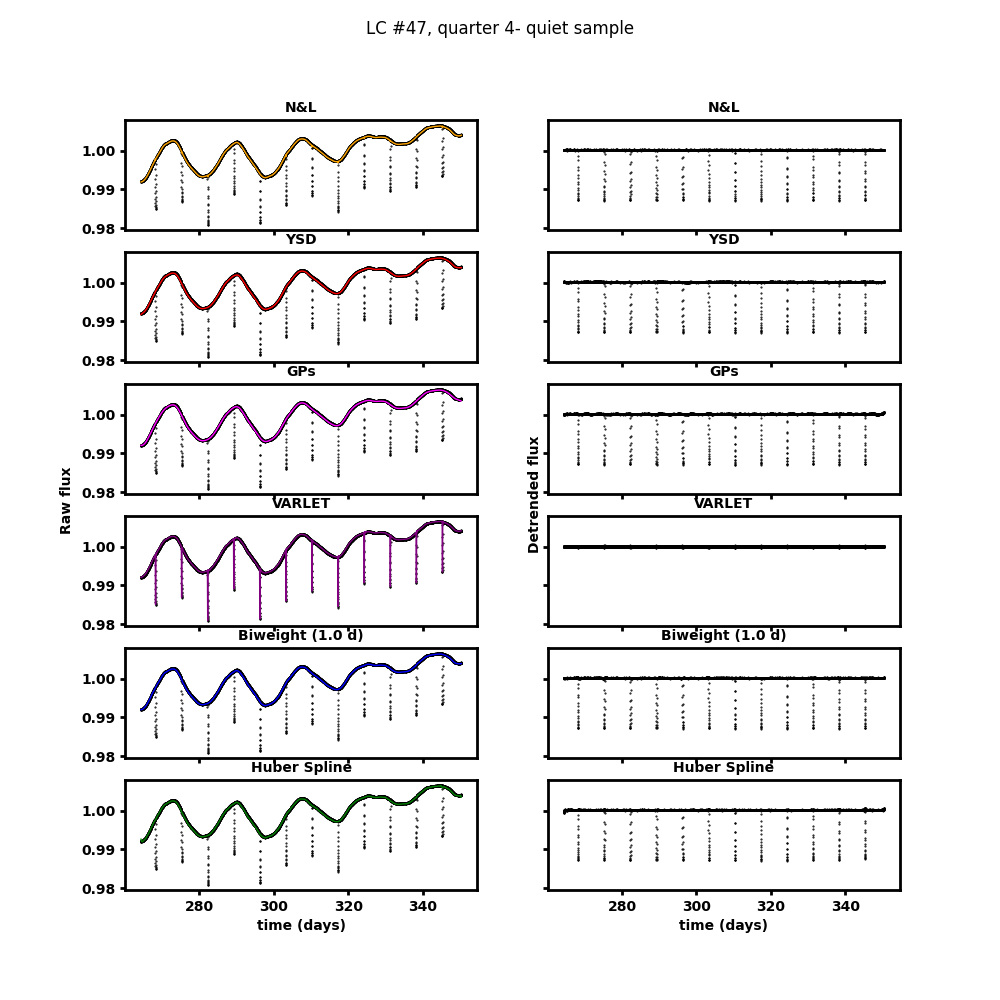}\hfil
\centering
\vspace{-25pt}
\caption{Filtering of quarter 4 of LC \#47 of the quiet sample with an injected large planet. The filtering models are overplotted in different colors and the resulting detrended flux is shown on the right panels.
}\label{fig: LC47q4_quiet} 
\centering
\end{figure}

\begin{figure}[H]
\centering
\includegraphics[width=.6\textwidth, center, right]{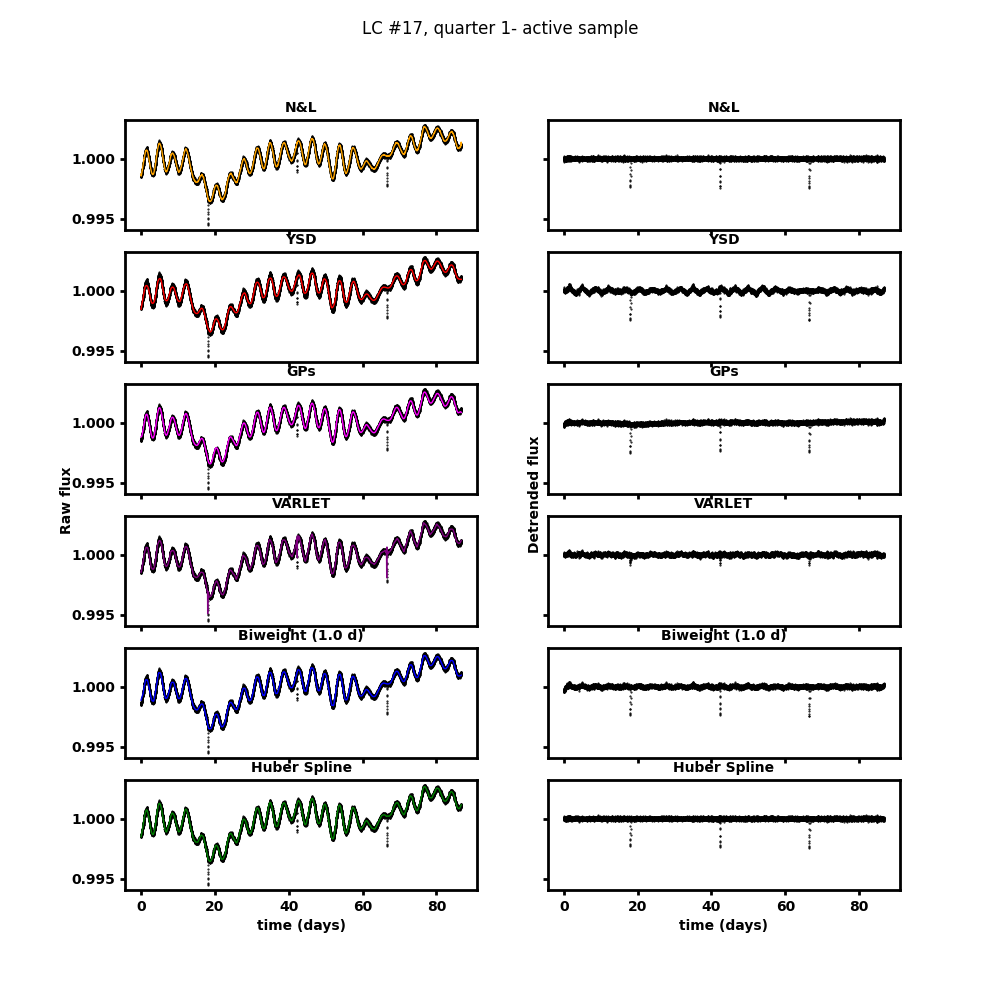}\hfil
\centering
\vspace{-15pt}
\caption{Filtering of quarter 1 of LC \#17 of the active sample with an injected large planet. The filtering models are overplotted in different colors and the resulting detrended flux is shown on the right panels.}\label{fig: LC17q1_active} 
\centering
\end{figure}

\begin{figure}[H]
\raggedright
\centering
\includegraphics[width=.6\textwidth, center, right]{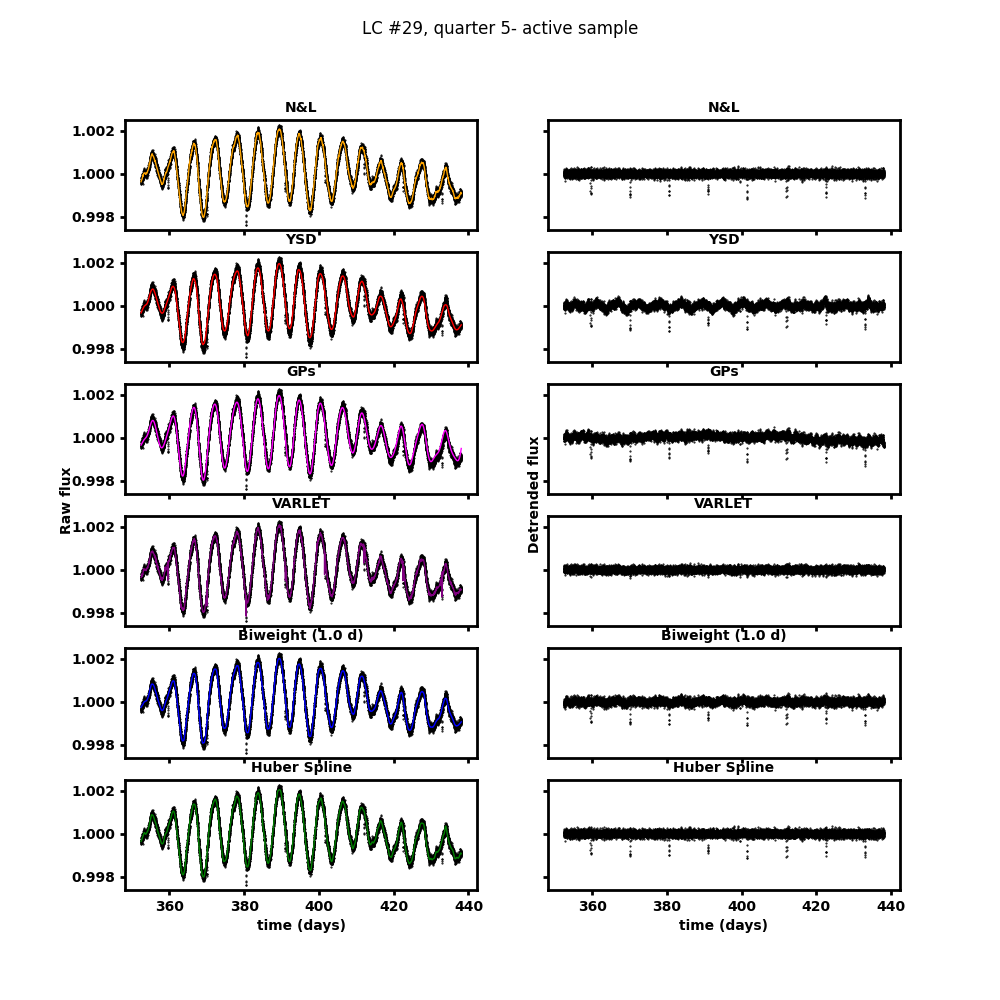}\hfil
\centering
\vspace{-15pt}
\caption{Filtering of quarter 5 of LC \#29 of the active sample with an injected large planet. The filtering models are overplotted in different colors and the resulting detrended flux is shown on the right panels.
}\label{fig: LC29q5_active} 
\centering
\end{figure}

\begin{figure}[H]
\centering
\includegraphics[width=.6\textwidth]{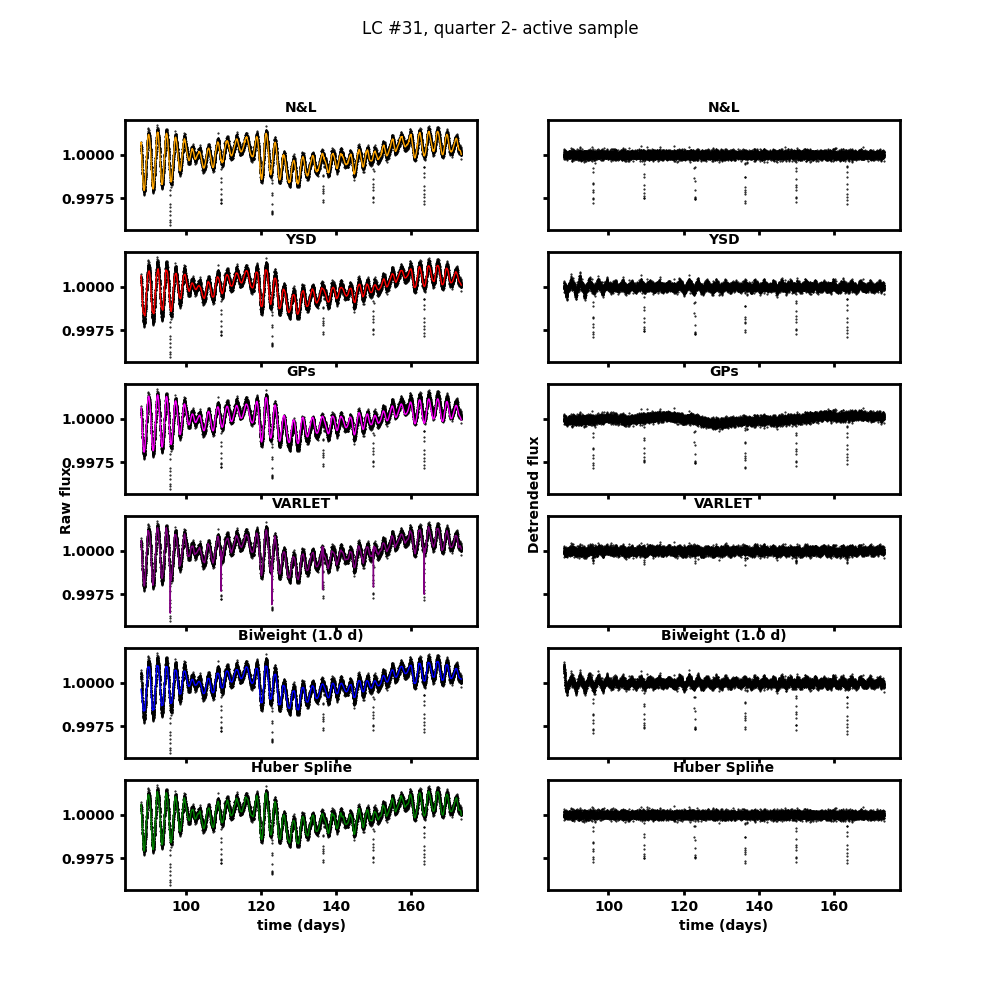}\hfil
\centering
\vspace{-25pt}
\caption{Filtering of quarter 2 of LC \#31 of the active sample with an injected large planet. The filtering models are overplotted in different colors and the resulting detrended flux is shown on the right panels.
}\label{fig: LC31q2_active} 
\centering
\end{figure}

\begin{figure}[H]
\centering
\includegraphics[width=.6\textwidth]{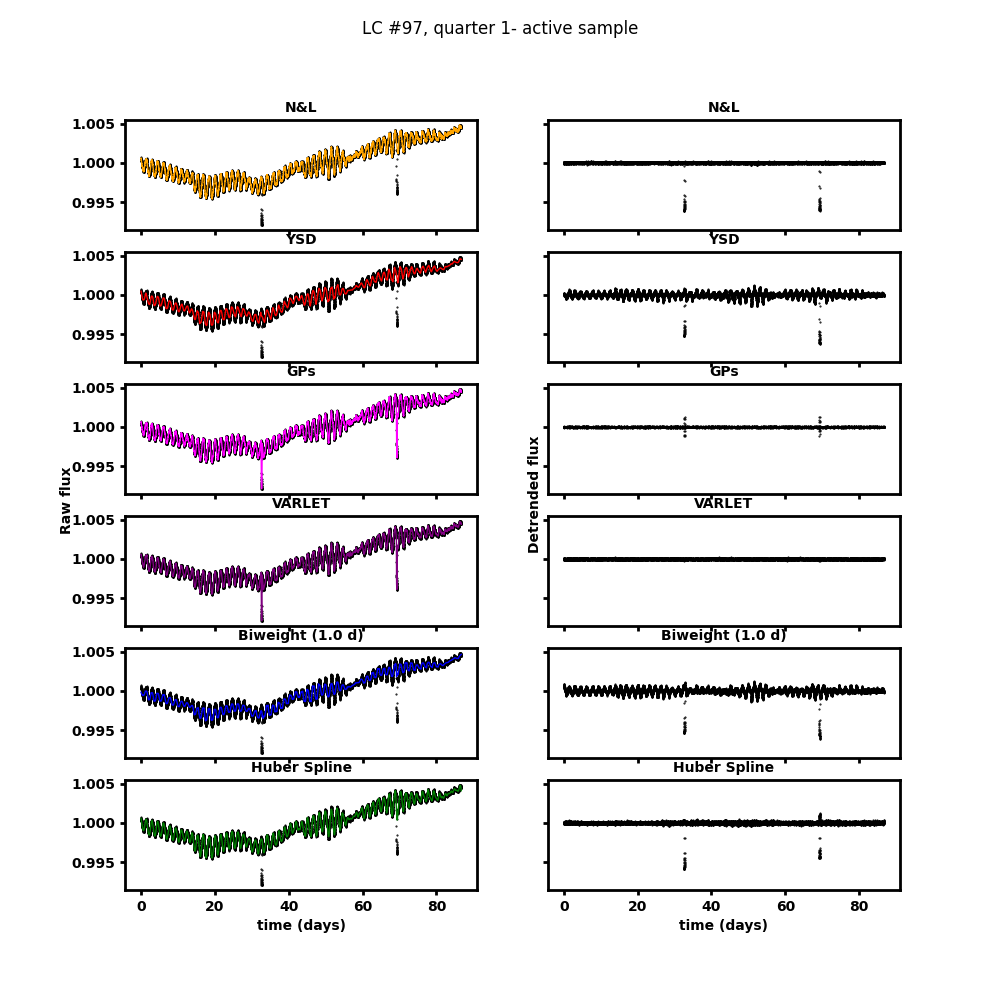}\hfil
\centering
\vspace{-25pt}
\caption{Filtering of quarter 1 of LC \#97 of the active sample with an injected large planet. The filtering models are overplotted in different colors and the resulting detrended flux is shown on the right panels.
}\label{fig: LC97q1_active} 
\centering
\end{figure}

The simulated light curves of both the quiet and active sample are provided as supplementary material. In Fig. \ref{fig: LC2q7_quiet}, \ref{fig: LC24q6_quiet}, \ref{fig: LC30q5_quiet}, and \ref{fig: LC47q4_quiet}, a selection of four simulated LCs from the quiet sample, before and after filtering is shown. A large planet (i.e., $R_\mathrm{p}/R_\star \geq 0.05$) is injected, as explained in Sect. \ref{sec: tests}. In these figures, six panels are displayed, representing the filtering performed by the selected algorithms. Specifically, in the left panels, we show the raw flux with the superimposed model, whereas in the right panels, we show the resulting filtered flux. Different colors are adopted for the model produced by different algorithms. In particular: Yellow for N\&L, red for YSD, fuchsia for GPs, violet for VARLET, blue for biweight, and green for the Huber spline. As a comparison, also a selection of 4 LCs from the active sample is shown in Fig. \ref{fig: LC17q1_active}, \ref{fig: LC29q5_active}, \ref{fig: LC31q2_active}, and \ref{fig: LC97q1_active}. Moreover, the full set of input parameters for the 100 simulated \textrm{PLATO} LCs of the quiet and active sample is reported in Table \ref{tab: LC param-quiet} and \ref{tab: LC param-active}, respectively.
\vspace{-30pt}
\vspace{-30pt}
\vspace{-30pt}
\vspace{-30pt}

    \begin{table*}
    \centering 
    \caption{Full set of input parameters for the 100 simulated \textrm{PLATO} light curves of the quiet sample. As explained in Sect. \ref{sec:PLATO LCs}, they were produced with PSLS 1.2 by the LSWG, with activity signals from the LCs simulated in \citet{Aigrain2015}, hereafter A15.}\label{tab: LC param-quiet} 
    \small 
    \begin{tabular}{c c c c c c c c c c c} \hline \hline 
    \#LC & ID in A15$^{(a)}$ & $V$ (mag) & activity level$^{(b)}$ & $P_\mathrm{rot}$ (days) & $T_\mathrm{eff}$ (K) & log$(g)$ (dex) & [Fe/H] & $\rho_\star$ (g/cm$^3$) & $u_1^{(c)}$ & $u_2^{(c)}$ \\ \hline 
0 & 913 & 9.75 & 0.32 & 31.26 & 5943.613 & 4.02 & 0.22 & 2.09 & 0.45 & 0.15\\ 
1 & 830 & 9.05 & 1.47 & 42.23 & 6648.757 & 4.32 & -0.11 & 2.02 & 0.19 & 0.57\\ 
2 & 955 & 9.94 & 1.11 & 27.83 & 5512.195 & 4.50 & 0.05 & 2.72 & 1.51 & -0.59\\ 
3 & 385 & 8.08 & 0.45 & 35.18 & 6648.757 & 4.32 & -0.11 & 2.39 & 1.52 & -0.58\\ 
4 & 285 & 9.74 & 0.71 & 50.19 & 5512.195 & 4.50 & 0.05 & 2.02 & 0.11 & 0.35\\ 
5 & 87 & 10.55 & 0.65 & 16.28 & 5943.613 & 4.02 & 0.22 & 0.83 & 0.03 & 0.54\\ 
6 & 720 & 8.39 & 0.59 & 56.31 & 5512.195 & 4.50 & 0.05 & 1.79 & 1.25 & -0.49\\ 
7 & 294 & 8.63 & 0.6 & 46.68 & 6516.527 & 4.02 & 0.10 & 2.30 & 0.04 & 0.37\\ 
8 & 430 & 9.45 & 0.44 & 22.01 & 5943.613 & 4.02 & 0.22 & 1.45 & 1.85 & -0.86\\ 
9 & 790 & 8.96 & 0.65 & 12.2 & 5512.195 & 4.50 & 0.05 & 1.35 & 0.11 & 0.04\\ 
10 & 581 & 8.83 & 0.35 & 1.65 & 6648.757 & 4.32 & -0.11 & 2.76 & 0.34 & 0.27\\ 
11 & 377 & 10.71 & 0.69 & 18.83 & 5943.613 & 4.02 & 0.22 & 2.15 & 1.14 & -0.20\\ 
12 & 117 & 9.57 & 1.86 & 10.4 & 5943.613 & 4.02 & 0.22 & 2.14 & 1.02 & -0.25\\ 
13 & 132 & 10.28 & 0.48 & 38.53 & 5943.613 & 4.02 & 0.22 & 2.81 & 0.45 & 0.48\\ 
14 & 399 & 9.09 & 0.66 & 39.83 & 6580.811 & 3.79 & -0.01 & 2.27 & 0.53 & -0.22\\ 
15 & 479 & 8.98 & 0.51 & 12.44 & 5512.195 & 4.50 & 0.05 & 1.15 & 0.28 & 0.39\\ 
16 & 960 & 10.93 & 0.39 & 25.21 & 6648.757 & 4.32 & -0.11 & 2.50 & 1.02 & -0.49\\ 
17 & 731 & 10.52 & 1.48 & 29.05 & 6580.811 & 3.79 & -0.01 & 2.69 & 1.13 & -0.17\\ 
18 & 424 & 10.49 & 3.03 & 68.37 & 6580.811 & 3.79 & -0.01 & 0.92 & 0.19 & -0.03\\ 
19 & 137 & 10.06 & 0.7 & 24.17 & 6516.527 & 4.02 & 0.10 & 1.64 & 1.80 & -0.87\\ 
20 & 49 & 10.58 & 0.79 & 36.6 & 6580.811 & 3.79 & -0.01 & 2.63 & 0.63 & 0.20\\ 
21 & 611 & 8.66 & 2.76 & 27.53 & 6516.527 & 4.02 & 0.10 & 1.86 & 0.56 & 0.35\\ 
22 & 787 & 8.78 & 1.24 & 2.4 & 5512.195 & 4.50 & 0.05 & 1.87 & 0.20 & 0.01\\ 
23 & 170 & 10.97 & 1.29 & 49.17 & 5512.195 & 4.50 & 0.05 & 1.62 & 0.65 & 0.08\\ 
24 & 914 & 10.33 & 0.32 & 30.26 & 5943.613 & 4.02 & 0.22 & 1.47 & 0.34 & 0.43\\ 
25 & 406 & 8.97 & 0.8 & 17.16 & 5943.613 & 4.02 & 0.22 & 2.76 & 1.27 & -0.33\\ 
26 & 954 & 9.36 & 0.36 & 19.8 & 5512.195 & 4.50 & 0.05 & 0.86 & 0.54 & 0.44\\ 
27 & 450 & 10.83 & 0.66 & 29.79 & 6580.811 & 3.79 & -0.01 & 1.97 & 0.84 & 0.03\\ 
28 & 416 & 8.61 & 2.71 & 30.1 & 6580.811 & 3.79 & -0.01 & 2.97 & 0.65 & -0.05\\ 
29 & 986 & 9.15 & 1.66 & 36.73 & 5512.195 & 4.50 & 0.05 & 1.62 & 1.50 & -0.67\\ 
30 & 442 & 9.58 & 2.48 & 18.56 & 5512.195 & 4.50 & 0.05 & 1.83 & 0.88 & -0.42\\ 
31 & 594 & 8.42 & 0.89 & 13.04 & 6580.811 & 3.79 & -0.01 & 2.09 & 1.54 & -0.61\\ 
32 & 670 & 8.56 & 0.62 & 10.9 & 5943.613 & 4.02 & 0.22 & 2.60 & 0.11 & 0.26\\ 
33 & 13 & 9.42 & 2.63 & 46.51 & 5512.195 & 4.50 & 0.05 & 1.00 & 0.99 & -0.23\\ 
34 & 112 & 8.77 & 0.39 & 13.57 & 5512.195 & 4.50 & 0.05 & 2.45 & 0.29 & 0.47\\ 
35 & 408 & 10.46 & 1.03 & 40.55 & 5943.613 & 4.02 & 0.22 & 2.79 & 0.35 & 0.56\\ 
36 & 628 & 10.28 & 1.17 & 1.65 & 6516.527 & 4.02 & 0.10 & 1.12 & 0.06 & 0.29\\ 
37 & 523 & 10.05 & 1.63 & 10.24 & 6516.527 & 4.02 & 0.10 & 1.63 & 0.20 & 0.11\\ 
38 & 554 & 10.03 & 1.32 & 32.32 & 6648.757 & 4.32 & -0.11 & 2.06 & 1.43 & -0.64\\ 
39 & 751 & 10.88 & 1.65 & 30.77 & 6516.527 & 4.02 & 0.10 & 1.76 & 1.22 & -0.61\\ 
40 & 251 & 9.46 & 2.83 & 12.99 & 5943.613 & 4.02 & 0.22 & 1.83 & 0.28 & -0.02\\ 
41 & 960 & 8.1 & 0.39 & 25.21 & 6516.527 & 4.02 & 0.10 & 1.26 & 0.08 & 0.83\\ 
42 & 277 & 8.7 & 1.0 & 1.37 & 6580.811 & 3.79 & -0.01 & 1.97 & 1.17 & -0.44\\ 
43 & 16 & 9.25 & 2.59 & 18.24 & 6580.811 & 3.79 & -0.01 & 1.75 & 0.90 & -0.19\\ 
44 & 446 & 9.71 & 0.4 & 19.16 & 6580.811 & 3.79 & -0.01 & 2.12 & 0.36 & 0.53\\ 
45 & 191 & 9.63 & 1.0 & 21.54 & 6516.527 & 4.02 & 0.10 & 0.81 & 0.20 & 0.69\\ 
46 & 167 & 9.82 & 1.33 & 27.41 & 6580.811 & 3.79 & -0.01 & 1.56 & 0.99 & -0.02\\ 
47 & 784 & 8.24 & 3.06 & 17.49 & 6648.757 & 4.32 & -0.11 & 2.10 & 1.75 & -0.82\\ 
48 & 861 & 10.83 & 0.4 & 11.96 & 6516.527 & 4.02 & 0.10 & 2.84 & 0.66 & 0.28\\ 
49 & 384 & 9.32 & 0.95 & 18.85 & 5512.195 & 4.50 & 0.05 & 2.75 & 0.08 & 0.20\\ 
50 & 931 & 8.28 & 1.44 & 18.96 & 6580.811 & 3.79 & -0.01 & 2.86 & 0.28 & 0.03\\ 
51 & 785 & 8.82 & 2.77 & 3.94 & 6580.811 & 3.79 & -0.01 & 1.40 & 0.85 & 0.10\\ 
52 & 484 & 9.01 & 1.71 & 53.07 & 5512.195 & 4.50 & 0.05 & 3.00 & 0.01 & 0.37\\ 
53 & 237 & 9.55 & 2.99 & 43.95 & 6648.757 & 4.32 & -0.11 & 2.76 & 0.38 & -0.08\\ 
54 & 232 & 8.39 & 0.53 & 33.86 & 5943.613 & 4.02 & 0.22 & 1.38 & 0.08 & 0.49\\ 
55 & 516 & 9.5 & 0.95 & 19.21 & 5943.613 & 4.02 & 0.22 & 1.21 & 0.01 & 0.66\\ 
56 & 778 & 9.63 & 1.85 & 25.62 & 6516.527 & 4.02 & 0.10 & 2.84 & 0.97 & -0.45\\ 
57 & 901 & 10.53 & 0.35 & 15.78 & 6648.757 & 4.32 & -0.11 & 0.84 & 0.97 & -0.10\\ 
58 & 202 & 10.9 & 0.68 & 46.41 & 5512.195 & 4.50 & 0.05 & 1.03 & 0.99 & -0.48\\ 
59 & 755 & 8.01 & 2.98 & 3.33 & 6516.527 & 4.02 & 0.10 & 2.77 & 0.68 & -0.03\\ 
60 & 644 & 9.77 & 1.08 & 17.12 & 5943.613 & 4.02 & 0.22 & 2.31 & 0.66 & 0.21\\ 
61 & 157 & 10.17 & 0.33 & 22.94 & 5943.613 & 4.02 & 0.22 & 1.09 & 1.45 & -0.58\\ 
62 & 914 & 8.77 & 0.32 & 30.26 & 5512.195 & 4.50 & 0.05 & 1.78 & 0.42 & -0.16\\ 
63 & 140 & 10.05 & 3.06 & 18.29 & 5512.195 & 4.50 & 0.05 & 2.94 & 0.43 & 0.17\\ 
64 & 350 & 10.11 & 0.72 & 16.83 & 5943.613 & 4.02 & 0.22 & 2.31 & 0.31 & 0.42\\ 
65 & 741 & 9.07 & 0.94 & 63.1 & 5512.195 & 4.50 & 0.05 & 0.97 & 0.25 & 0.04\\ 
66 & 960 & 9.77 & 0.39 & 25.21 & 5943.613 & 4.02 & 0.22 & 0.99 & 0.48 & 0.35\\ 
\end{tabular}
\end{table*} 
\begin{table*}
    \centering 
    \small 
    \begin{tabular}{c c c c c c c c c c c} 
    \#LC & ID in A15$^{(a)}$ & $V$ (mag) & activity level$^{(b)}$ & $P_\mathrm{rot}$ (days) & $T_\mathrm{eff}$ (K) & log$(g)$ (dex) & [Fe/H] &$\rho_\star$ (g/cm$^3$) & $u_1^{(c)}$ & $u_2^{(c)}$  \\ \hline
67 & 969 & 10.94 & 1.25 & 26.24 & 6648.757 & 4.32 & -0.11 & 1.52 & 1.04 & -0.26\\ 
68 & 532 & 10.52 & 1.35 & 13.63 & 6516.527 & 4.02 & 0.10 & 2.41 & 0.24 & 0.75\\ 
69 & 710 & 10.03 & 2.07 & 22.41 & 5512.195 & 4.50 & 0.05 & 1.34 & 0.18 & 0.45\\ 
70 & 448 & 9.84 & 0.71 & 45.65 & 5512.195 & 4.50 & 0.05 & 1.38 & 1.00 & -0.22\\ 
71 & 996 & 10.37 & 2.22 & 5.72 & 6516.527 & 4.02 & 0.10 & 1.31 & 0.08 & 0.15\\ 
72 & 353 & 10.49 & 2.56 & 30.52 & 5512.195 & 4.50 & 0.05 & 1.75 & 0.20 & 0.46\\ 
73 & 200 & 9.2 & 1.39 & 11.68 & 6648.757 & 4.32 & -0.11 & 2.47 & 0.10 & 0.54\\ 
74 & 300 & 10.83 & 0.52 & 26.01 & 6648.757 & 4.32 & -0.11 & 1.14 & 0.85 & 0.01\\ 
75 & 456 & 9.14 & 0.55 & 30.74 & 6516.527 & 4.02 & 0.10 & 1.72 & 1.08 & -0.39\\ 
76 & 403 & 10.95 & 1.15 & 38.55 & 6648.757 & 4.32 & -0.11 & 2.10 & 0.22 & 0.52\\ 
77 & 253 & 10.25 & 2.04 & 39.75 & 5512.195 & 4.50 & 0.05 & 2.38 & 0.29 & -0.04\\ 
78 & 516 & 10.71 & 0.95 & 19.21 & 6516.527 & 4.02 & 0.10 & 2.29 & 0.35 & 0.01\\ 
79 & 482 & 8.18 & 0.59 & 22.56 & 6580.811 & 3.79 & -0.01 & 2.75 & 0.05 & 0.57\\ 
80 & 29 & 10.03 & 0.64 & 35.9 & 6580.811 & 3.79 & -0.01 & 2.08 & 0.02 & 0.51\\ 
81 & 489 & 9.31 & 0.72 & 20.44 & 5943.613 & 4.02 & 0.22 & 2.22 & 0.42 & 0.07\\ 
82 & 257 & 8.96 & 0.91 & 36.94 & 6648.757 & 4.32 & -0.11 & 2.24 & 0.66 & 0.06\\ 
83 & 321 & 10.17 & 2.43 & 23.2 & 5512.195 & 4.50 & 0.05 & 2.68 & 0.74 & 0.03\\ 
84 & 547 & 9.53 & 0.67 & 4.78 & 5512.195 & 4.50 & 0.05 & 2.00 & 0.44 & 0.19\\ 
85 & 109 & 10.52 & 0.52 & 37.27 & 5512.195 & 4.50 & 0.05 & 0.94 & 0.80 & 0.16\\ 
86 & 849 & 8.0 & 0.62 & 45.61 & 5512.195 & 4.50 & 0.05 & 1.17 & 0.26 & 0.71\\ 
87 & 989 & 10.85 & 0.6 & 24.41 & 6516.527 & 4.02 & 0.10 & 2.61 & 0.84 & -0.33\\ 
88 & 502 & 9.21 & 0.5 & 44.51 & 6516.527 & 4.02 & 0.10 & 2.36 & 0.11 & 0.39\\ 
89 & 69 & 9.85 & 0.84 & 36.08 & 5943.613 & 4.02 & 0.22 & 1.27 & 0.10 & 0.89\\ 
90 & 668 & 9.29 & 0.52 & 52.15 & 5943.613 & 4.02 & 0.22 & 2.62 & 0.12 & 0.06\\ 
91 & 132 & 8.86 & 0.48 & 38.53 & 5512.195 & 4.50 & 0.05 & 2.02 & 0.03 & 0.68\\ 
92 & 698 & 10.18 & 1.54 & 16.09 & 5943.613 & 4.02 & 0.22 & 0.84 & 0.90 & -0.05\\ 
93 & 248 & 8.65 & 1.89 & 21.07 & 6648.757 & 4.32 & -0.11 & 1.96 & 0.35 & 0.33\\ 
94 & 550 & 8.5 & 0.36 & 52.63 & 6516.527 & 4.02 & 0.10 & 2.43 & 0.13 & -0.03\\ 
95 & 716 & 9.37 & 0.93 & 12.26 & 6648.757 & 4.32 & -0.11 & 2.86 & 0.44 & 0.21\\ 
96 & 684 & 8.76 & 2.39 & 31.97 & 6648.757 & 4.32 & -0.11 & 1.06 & 0.55 & 0.14\\ 
97 & 213 & 9.36 & 1.73 & 11.22 & 6580.811 & 3.79 & -0.01 & 1.62 & 0.58 & -0.22\\ 
98 & 382 & 8.22 & 2.83 & 42.31 & 6648.757 & 4.32 & -0.11 & 2.42 & 0.18 & 0.82\\ 
99 & 76 & 8.98 & 0.42 & 21.05 & 6580.811 & 3.79 & -0.01 & 2.40 & 1.49 & -0.68\\     \hline 
    \end{tabular}\\
    \raggedright  
\textbf{Notes:}\\ 
$^{(a)}$ "ID in A15" is the stellar identifier in A15.\\
$^{(b)}$ activity level is relative to the solar.\\
$^{(c)}$ $u_1$ and $u_2$ are the limb darkening coefficients, assuming a quadratic limb darkening law (\citealt{Kopal1950}).
    \end{table*}

    \begin{table}
    \centering 
    \caption{Full set of input parameters for the 100 simulated \textrm{PLATO} light curves of the active sample. The following parameters are assumed to be the same of the quiet sample: effective temperature ($T_\mathrm{eff}$), gravity (log($g$)), metallicity ([Fe/H]), density ($\rho_\star$), and limb darkening parameters ($u_1$ and $u_2$).}\label{tab: LC param-active} 
    \small 
    \begin{tabular}{c c c c c} \hline \hline 
    \# LC & ID in A15 & $V$ (mag) & activity level & $P_\mathrm{rot}$ (days) \\ \hline 
0 & 421 & 9.75 & 2.39 & 6.81\\ 
1 & 908 & 9.05 & 0.40 & 6.41\\ 
2 & 0 & 9.94 & 0.39 & 1.51\\ 
3 & 277 & 8.08 & 1.00 & 1.37\\ 
4 & 573 & 9.74 & 1.65 & 2.80\\ 
5 & 364 & 10.55 & 1.05 & 5.66\\ 
6 & 889 & 8.39 & 0.44 & 7.34\\ 
7 & 459 & 8.63 & 1.71 & 2.17\\ 
8 & 742 & 9.45 & 1.12 & 9.42\\ 
9 & 320 & 8.96 & 0.88 & 4.66\\ 
10 & 797 & 8.83 & 0.42 & 8.27\\ 
11 & 742 & 10.71 & 1.12 & 9.42\\ 
12 & 514 & 9.57 & 0.55 & 7.97\\ 
13 & 547 & 10.28 & 0.67 & 4.78\\ 
14 & 676 & 9.09 & 0.96 & 2.28\\ 
15 & 418 & 8.98 & 0.35 & 3.83\\ 
16 & 150 & 10.93 & 1.15 & 1.94\\ 
17 & 706 & 10.52 & 2.57 & 3.67\\ 
18 & 581 & 10.49 & 0.35 & 1.65\\ 
19 & 609 & 10.06 & 2.13 & 2.69\\ 
20 & 67 & 10.58 & 0.93 & 1.32\\ 
21 & 547 & 8.66 & 0.67 & 4.78\\ 
22 & 742 & 8.78 & 1.12 & 9.42\\ 
23 & 266 & 10.97 & 2.04 & 4.51\\ 
24 & 393 & 10.33 & 0.46 & 9.89\\ 
25 & 118 & 8.97 & 1.46 & 2.27\\ 
26 & 436 & 9.36 & 0.92 & 7.61\\ 
27 & 891 & 10.83 & 0.44 & 8.56\\ 
28 & 36 & 8.61 & 0.82 & 2.04\\ 
29 & 996 & 9.15 & 2.22 & 5.72\\ 
30 & 233 & 9.58 & 0.35 & 4.49\\ 
31 & 719 & 8.42 & 2.30 & 2.44\\ 
32 & 787 & 8.56 & 1.24 & 2.40\\ 
33 & 750 & 9.42 & 0.45 & 5.61\\ 
34 & 638 & 8.77 & 0.33 & 8.53\\ 
35 & 649 & 10.46 & 0.41 & 7.75\\ 
36 & 728 & 10.28 & 0.61 & 7.86\\ 
37 & 390 & 10.05 & 2.40 & 3.17\\ 
38 & 36 & 10.03 & 0.82 & 2.04\\ 
39 & 750 & 10.88 & 0.45 & 5.61\\ 
40 & 984 & 9.46 & 0.48 & 2.50\\ 
41 & 547 & 8.10 & 0.67 & 4.78\\ 
42 & 996 & 8.70 & 2.22 & 5.72\\ 
43 & 854 & 9.25 & 1.45 & 1.24\\ 
44 & 755 & 9.71 & 2.98 & 3.33\\ 
45 & 638 & 9.63 & 0.33 & 8.53\\ 
46 & 638 & 9.82 & 0.33 & 8.53\\ 
47 & 23 & 8.24 & 0.42 & 3.31\\ 
48 & 609 & 10.83 & 2.13 & 2.69\\ 
49 & 872 & 9.32 & 0.51 & 6.87\\ 
50 & 233 & 8.28 & 0.35 & 4.49\\ 
51 & 785 & 8.82 & 2.77 & 3.94\\ 
52 & 956 & 9.01 & 0.59 & 7.45\\ 
53 & 1 & 9.55 & 0.45 & 1.87\\ 
54 & 854 & 8.39 & 1.45 & 1.24\\ 
55 & 961 & 9.50 & 0.50 & 8.29\\ 
56 & 369 & 9.63 & 0.51 & 9.22\\ 
57 & 277 & 10.53 & 1.00 & 1.37\\ 
58 & 753 & 10.90 & 0.78 & 1.90\\ 
59 & 785 & 8.01 & 2.77 & 3.94\\ 
60 & 996 & 9.77 & 2.22 & 5.72\\ 
61 & 390 & 10.17 & 2.40 & 3.17\\ 
62 & 744 & 8.77 & 2.69 & 3.12\\ 
 \end{tabular}
    \raggedright  
    \end{table}
 \begin{table}
    \centering 
    \small 
    \begin{tabular}{c c c c c}  
    \# LC & ID in A15 & $V$ (mag) & activity level & $P_\mathrm{rot}$ (days) \\ \hline 
63 & 427 & 10.05 & 1.52 & 2.68\\ 
64 & 753 & 10.11 & 0.78 & 1.90\\ 
65 & 909 & 9.07 & 0.32 & 1.58\\ 
66 & 943 & 9.77 & 0.36 & 6.72\\ 
67 & 150 & 10.94 & 1.15 & 1.94\\ 
68 & 889 & 10.52 & 0.44 & 7.34\\ 
69 & 678 & 10.03 & 0.81 & 4.94\\ 
70 & 116 & 9.84 & 1.18 & 9.19\\ 
71 & 67 & 10.37 & 0.93 & 1.32\\ 
72 & 940 & 10.49 & 2.90 & 6.58\\ 
73 & 678 & 9.20 & 0.81 & 4.94\\ 
74 & 266 & 10.83 & 2.04 & 4.51\\ 
75 & 854 & 9.14 & 1.45 & 1.24\\ 
76 & 597 & 10.95 & 2.09 & 5.89\\ 
77 & 649 & 10.25 & 0.41 & 7.75\\ 
78 & 649 & 10.71 & 0.41 & 7.75\\ 
79 & 673 & 8.18 & 0.69 & 2.13\\ 
80 & 181 & 10.03 & 0.49 & 2.08\\ 
81 & 118 & 9.31 & 1.46 & 2.27\\ 
82 & 277 & 8.96 & 1.00 & 1.37\\ 
83 & 436 & 10.17 & 0.92 & 7.61\\ 
84 & 797 & 9.53 & 0.42 & 8.27\\ 
85 & 841 & 10.52 & 2.41 & 8.17\\ 
86 & 956 & 8.00 & 0.59 & 7.45\\ 
87 & 706 & 10.85 & 2.57 & 3.67\\ 
88 & 823 & 9.21 & 0.33 & 3.22\\ 
89 & 622 & 9.85 & 0.92 & 8.08\\ 
90 & 445 & 9.29 & 0.89 & 6.29\\ 
91 & 118 & 8.86 & 1.46 & 2.27\\ 
92 & 943 & 10.18 & 0.36 & 6.72\\ 
93 & 940 & 8.65 & 2.90 & 6.58\\ 
94 & 785 & 8.50 & 2.77 & 3.94\\ 
95 & 823 & 9.37 & 0.33 & 3.22\\ 
96 & 67 & 8.76 & 0.93 & 1.32\\ 
97 & 563 & 9.36 & 2.64 & 1.56\\ 
98 & 909 & 8.22 & 0.32 & 1.58\\ 
99 & 984 & 8.98 & 0.48 & 2.50\\ \hline 
     \end{tabular}
    \raggedright  
    \end{table}
\clearpage
\section{Comparison of running times}\label{app: times}
In Table \ref{tab: times}, we report the running times of the filtering algorithms analyzed in this work, we computed over 20 \textrm{PLATO} simulated LCs of the active sample encompassing one single quarter (including 12486 data points). The analysis was carried out on an outdated desktop machine hosting two Intel(R) Xeon(R) E5-2620 at 2.00GHz CPUs for a total of 24 threads and a 2TB hard drive spinning at 5400 RPM.
We did not make any attempt to optimize the (extremely heterogeneous) preexisting implementations of the algorithms or to allow for multiprocessing analysis if not already implemented, as this task was beyond the scope of this work. For these reasons, we did not consider the execution time a relevant factor in our comparisons.
    \begin{table}[H]
    \centering 
    \caption{Comparison of running times (in seconds) of the six tested filtering algorithms for the first quarter of 20 \textrm{PLATO} light curves of the active sample with an injected large planet.}\label{tab: times} 
    \small 
    \begin{tabular}{c c c c c c c} \hline \hline 
    \#LC & N\&L & YSD & GPs & VARLET & Biweight (1.0 d) & Huber spline \\ \hline 
0 & 213.01 s & 6.15 s & 266.76 s & 11.26 s & 0.46 s & 3.27 s \\
1 & 217.30 s & 6.09 s & 107.33 s & 11.07 s & 0.36 s & 3.38 s \\
2 & 200.43 s & 6.05 s & 107.93 s & 11.24 s & 0.37 s & 3.37 s\\
3 & 197.82 s & 6.05 s & 110.24 s & 11.28 s & 0.36 s & 2.97 s \\
4 & 193.07 s & 6.08 s & 107.71 s & 11.06 s & 0.37 s & 2.96 s\\
5 & 209.88 s & 6.09 s & 108.04 s & 11.24 s & 0.37 s & 3.45 s \\
6 & 209.04 s & 6.11 s & 111.11 s & 11.23 s & 0.36 s & 3.18 s\\
7 & 203.67 s & 6.03 s & 110.08 s & 11.12 s & 0.37 s & 3.42 s\\
8 & 206.90 s & 6.01 s & 110.96 s & 11.21 s & 0.35 s & 3.37 s\\
9 & 199.69 s & 6.12 s & 111.08 s & 11.11 s & 0.36 s & 3.39 s\\
10 & 220.26 s & 6.02 s & 110.86 s & 11.08 s & 0.36 s & 3.51 s\\
11 & 196.25 s & 6.15 s & 110.83 s & 11.17 s & 0.35 s & 3.13 s\\
12 & 205.31 s & 6.15 s & 111.83 s & 11.10 s & 0.34 s & 3.17 s\\
13 & 200.09 s & 6.13 s & 113.65 s & 11.02 s & 0.35 s & 3.28 s\\
14 & 206.92 s & 6.13 s & 111.71 s & 11.03 s & 0.36 s & 3.53 s\\
15 & 213.02 s & 6.07 s & 112.98 s & 11.31 s & 0.37 s & 3.43 s\\
16 & 198.87 s & 6.17 s & 113.63 s & 11.09 s & 0.36 s & 3.21 s\\
17 & 227.97 s & 6.03 s & 112.00 s & 11.18 s & 0.37 s & 3.36 s\\
18 & 218.49 s & 6.12 s & 114.87 s & 11.04 s & 0.36 s & 3.37 s\\
19 & 209.87 s & 6.00 s & 116.54 s & 11.28 s & 0.37 s & 3.21 s \\ \hline 
Mean & 207.39 s & 6.08 s & 119.01 s & 11.16 s & 0.37 s & 3.29 s \\ \hline 
\end{tabular}
\end{table}

%
%




\end{appendix}
\end{document}